\documentclass[twoside]{article}
\usepackage{amsfonts,amssymb,amsbsy,textcomp,marvosym,amsmath,caption,threeparttable,amsthm,subfigure}
\usepackage{eurosym,mathrsfs,fancyhdr,CJK,multicol,graphics,indentfirst,color,bm,upgreek,booktabs,graphicx}
\usepackage{multirow}
\usepackage{xcolor}
\def\footnoterule{\kern 1mm \hrule width 10cm \kern 2mm}

\allowdisplaybreaks
\catcode`@=11
\def\title#1{\vspace{3mm}\begin{flushleft}\vglue-.1cm\Large\bf\boldmath\protect\baselineskip=18pt plus.2pt minus.1pt #1
\end{flushleft}\vspace{1mm} }
\def\author#1{\begin{flushleft}\normalsize #1\end{flushleft}\vspace*{-4pt} \vspace{3mm}}
\def\address#1#2{\begin{flushleft}\vglue-.35cm${}^{#1}$\small\it #2\vglue-.35cm\end{flushleft}\vspace{-2mm}\par}

\catcode`@=11
\def\section{\@startsection{section}{1}{\z@}%
 {-3ex \@plus -.3ex \@minus -.2ex}%
 {2.2ex \@plus.2ex}%
{\normalfont\normalsize\protect\baselineskip=14.5pt plus.2pt minus.2pt\bfseries}}
\def\subsection{\@startsection{subsection}{2}{\z@}%
 {-3ex\@plus -.2ex \@minus -.2ex}%
 {2ex \@plus.2ex}%
{\normalfont\normalsize\protect\baselineskip=12.5pt plus.2pt minus.2pt\bfseries}}
\def\subsubsection{\@startsection{subsubsection}{3}{\z@}%
 {-2.2ex\@plus -.21ex \@minus -.2ex}%
 {1.4ex \@plus.2ex}
{\normalfont\normalsize\protect\baselineskip=12pt plus.2pt minus.2pt\sl}}

\newcommand{\yyj}[1]{{\color{black}#1}}

\begin{document}
\begin{CJK*}{GBK}{song}
\thispagestyle{empty}
\vspace*{-13mm}
\vspace*{2mm}

\title{A Revisit of Shape Editing Techniques: from the Geometric to the Neural Viewpoint}

\author{Yu-Jie Yuan$^{1,2}$, Yu-Kun Lai$^{3}$, Tong Wu$^{1,2}$, Lin Gao$^{1,2,*}$, and Ligang Liu$^{4}$}

\address{1}{Beijing Key Laboratory of Mobile Computing and Pervasive Device, Institute of Computing Technology, Chinese Academy of Sciences}
\address{2}{University of Chinese Academy of Sciences}
\address{3}{School of Computer Science and Informatics, Cardiff University}
\address{4}{University of Science and Technology of China}

\let\thefootnote\relax\footnotetext{{}\\[-4mm] 
\indent\ $^*$Corresponding Author, E-mail: gaolin@ict.ac.cn
}

\noindent {\small\bf Abstract} \quad  {\small 
3D shape editing is widely used in a range of applications such as movie production, computer games and computer aided design. It is also a popular research topic in computer graphics and computer vision. In past decades, researchers have developed a series of editing methods to make the editing process faster, more robust, and more reliable. Traditionally, the deformed shape is determined by the optimal transformation and weights for an energy term. With increasing availability of 3D shapes on the Internet, data-driven methods were proposed to improve the editing results. More recently as the deep neural networks became popular, many deep learning based editing methods have been developed in this field\yyj{, which is naturally data-driven}. 
\yyj{We mainly survey recent research works from the geometric viewpoint to those emerging neural deformation techniques and categorize them into organic shape editing methods and man-made model editing methods.}
Both traditional methods and recent neural network based methods are reviewed.
}

\vspace*{3mm}


\vspace*{4mm}

\end{CJK*}
\baselineskip=18pt plus.2pt minus.2pt
\parskip=0pt plus.2pt minus0.2pt
\begin{multicols}{2}

\section{Introduction}
3D shapes are one of the most important types of objects in computer graphics and computer vision research. Editing or interactive deformation for 3D shapes provides an intuitive way to produce new shapes based on existing ones, which is fundamental for many applications. 
Methods for 3D shape editing are therefore one of the research hot spots. 
In recent years, deep learning has been widely used, and many research fields have developed new solutions based on deep learning, such as deep generation of 3D models~\cite{gao2019sdm, yang2020dsm}, 3D deep reconstruction~\cite{chen2019learning, xu2019disn}, deep neural network based 3D shape analysis methods~\cite{gao2019prs, shi2020symmetrynet}, etc. 3D models can be generally divided into two types, namely organic shapes and man-made models. Fig.1 shows some examples of these two types. Organic shapes such as human bodies, animals, etc. are often deformable, whereas man-made objects tend to comprise of a larger number of (near-)rigid components.
Different techniques are therefore needed to cope with these two types of shapes. Neural network based editing methods based on deep learning are also emerging, although they are still at relatively early stage and many open areas remain, which we will discuss later in the survey.

Early 3D model editing methods analyze the characteristics of the model itself, and strive to keep these characteristics unchanged during the deformation. For organic shapes, common examples include human bodies and animal shapes, which are articulated as shown in the left of Fig.1. It is possible to bind a skeleton inside the model. On the one hand, the editing of these models typically defines deformation energies to impose constraints on the deformation, such as volume-preserving deformation. On the other hand, by binding skeletons for these models, the user can manipulate the skeleton to drive the deformation of the shape. Skeleton-based deformation is often convenient and leads to good results. However, the binding of the skeleton is not only time-consuming, but also requires professional software and expertise. For man-made models, the main purpose of editing is modify the appearance, or geometric features of the models. For this purpose the topological structure of the model is usually a feature that needs to be maintained. 
Such kind of methods is referred to as structure-aware editing~\cite{mitra2014structure}. The editing of man-made models is more complicated than the deformation of organic shapes, because organic shapes are typically manifold meshes, while man-made models are often non-manifold with more complex structures.

Surveys on other aspects of 3D models have recently been published, such as 3D deep generative models~\cite{chaudhuri2020learning}, 3D deep reconstruction~\cite{han2019image, jin20203d} and 3D deep representation~\cite{xiao2020survey}. However, for 3D shape editing/deformation,
existing surveys~\cite{botsch2008linear, gain2008survey} were published over a decade ago, which only cover deformation methods of 3D organic shapes. Methods for editing of man-made models are not reviewed in specialized surveys, and only discussed in loosely related courses~\cite{mitra2014structure,xu2016data}. The rapid development of deep learning in recent years has also led to the emergence and growth of neural network based deformation and editing methods. It is necessary to have an extensive
review to summarize the related research and discuss future directions. To this end, we present this survey, reviewing both traditional methods and methods based on deep neural networks, as well as methods applied to both organic shapes and man-made models. 

\yyj{The structure of this survey is as follows. We divide the editing methods according to different analysis views, namely geometry-based (Sec.~\ref{sec:geometry}) or traditional data-driven based (Sec.~\ref{sec:dataset}). Although the neural-based editing methods also learn from dataset, because it is a new direction that is currently being actively explored and often requires and benefits from a larger amount of data, we will introduce it separately from the traditional data-driven methods in Sec.~\ref{sec:neural}. For these three types, because organic shapes and man-made models have certain differences in representation, and their editing methods also have certain differences, we summarize them for organic shapes and man-made models separately. Skeleton-based and cage-based deformation rely on a handle, which often be named as \textit{proxy}, different from the model itself and usually require weighted interpolation of deformation on the skeleton or cage to obtain the transformation of the shape itself. They will be included in Sec.~\ref{sec:proxy}, which can also use the information from dataset. Finally, we will conclude with existing problems and discuss interesting future research directions (Sec.~\ref{sec:conclusion}). Fig.~\ref{fig:timeline} provides a timeline of representative shape editing methods for organic shapes and man-made models.}

\begin{center}
\includegraphics[width=0.45\linewidth]{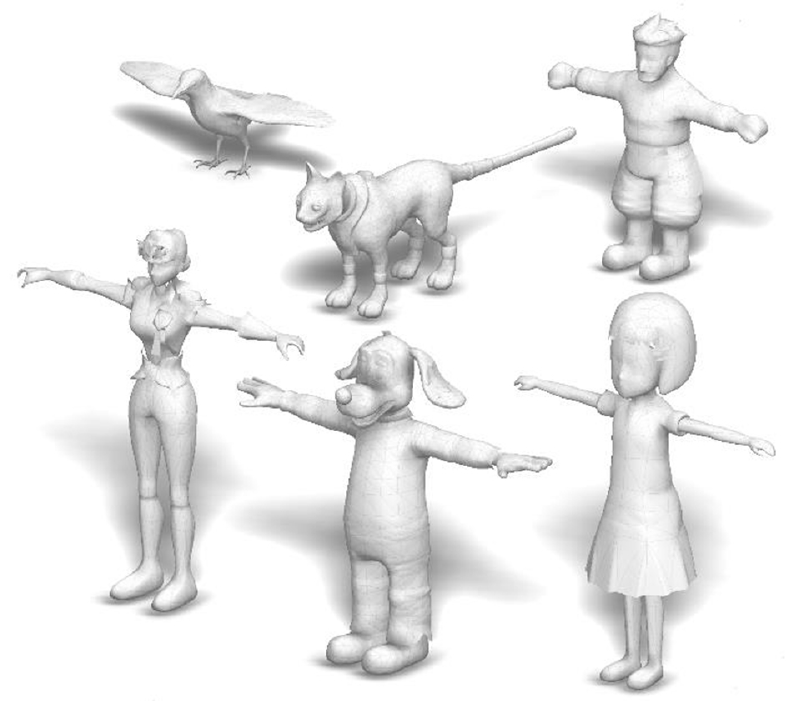}
\includegraphics[width=0.45\linewidth]{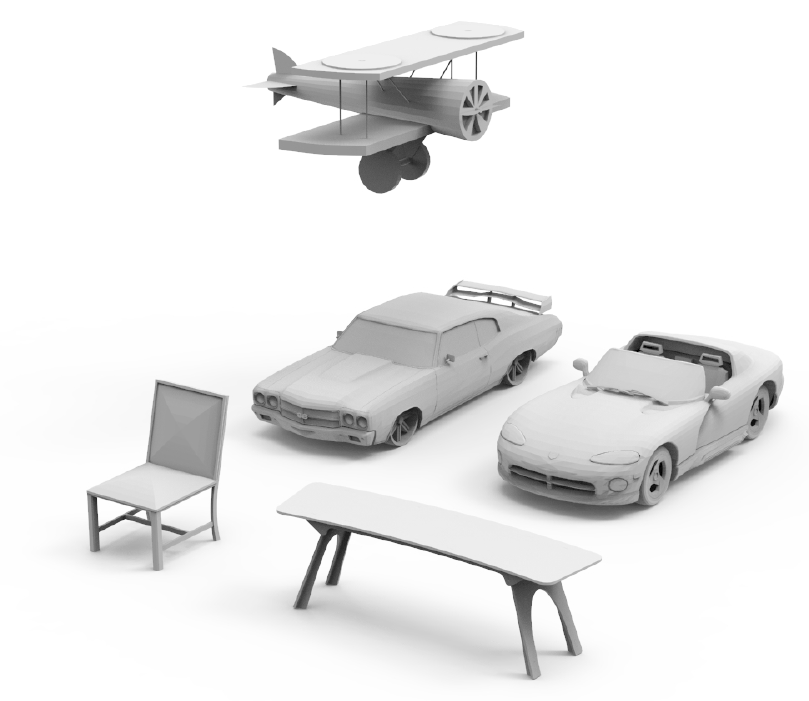}\\
\vspace{2mm}
\parbox[c]{8.3cm}{\footnotesize{Fig.1.~}Some examples of organic shapes (from~\cite{RigNet}) and man-made models (from~\cite{shapenet}). }%
\end{center}

\setcounter{figure}{1}
\begin{figure*}[t]
	\centering
	{
		\includegraphics[width=0.9\linewidth]{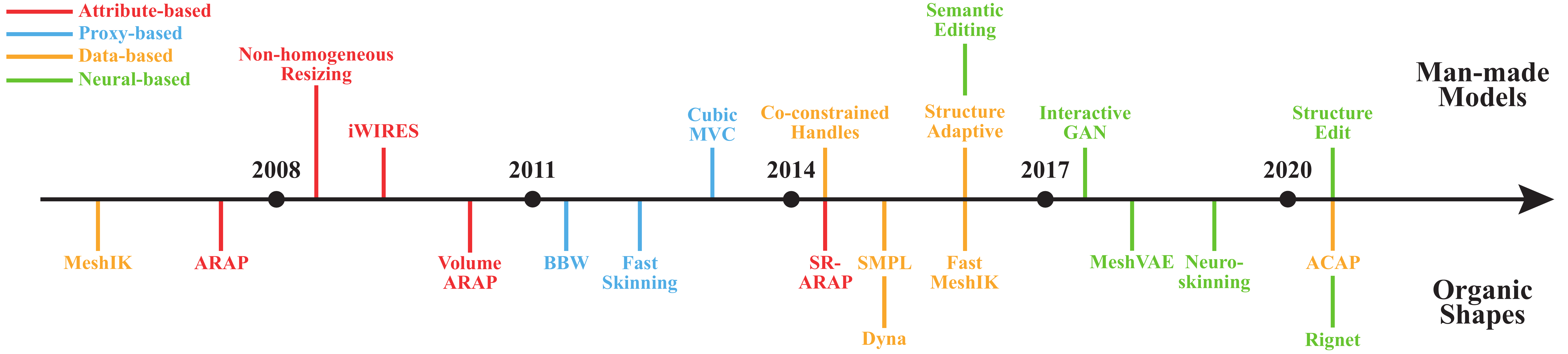}
	}
	\caption{The timeline of representative 3D shape editing methods for two types of 3D models.}
	\label{fig:timeline}
\end{figure*}

\section{Attribute-based Model Editing} \label{sec:geometry}
In this section, we discuss analyzing various attributes of the surface model, including geometry characteristics and semantic attributes, to define different constraints which then guide the model editing.

Organic shapes, a.k.a. deformable models, usually refer to models that are non-rigidly deformable. Human bodies and animal shapes are common examples. The editing of 3D organic shapes mostly uses interactive deformation. In these methods, organic shapes are often represented as triangular meshes. The deformation of organic shapes mainly strives to maintain the geometric details of the original shape and produce natural and reasonable results. The early deformation methods mainly analyze the geometry of the shape and define the constraints for the deformation. We summarize those methods as \emph{geometry-based mesh deformation} methods. 

The editing of man-made models will be relatively more complicated and difficult, compared to the deformation of organic shapes. On the one hand, the man-made models have different shapes and complex topological structures. On the other hand, the meshes of man-made models are generally not regular and consistent. This has certain obstacles to the direct application of some deformation methods of organic shapes. To achieve the purpose of editing, one should pose some constraints on 3D man-made models to ensure plausible results. One way to obtain those constraints is maintaining the structural relations between different components of the model, which is semantic knowledge. We summarize these \emph{semantic constraints for man-made models} in Sec.~\ref{sec:semantic}.

\subsection{Geometry-based Mesh Deformation} \label{sec:geometry}
Geometry-based deformation methods typically define energy functions which transform the deformation problem into a constrained optimization problem. The constraints are generally provided by the user by specifying control handles and their positions. Early research works were all around simulating the elastic deformation of objects. Terzopoulos et al.~\cite{terzopoulos1987elastically} proposed the classical elastic energy or the so-called \textit{shell energy} \yyj{which measuring stretching and bending by the change of the first and second fundamental forms} and optimize the energy to obtain deformation results. \yyj{Two follow-up works~\cite{celniker1991deformable, welch1992variational} propose to simplify the energy by replacing the first and second fundamental forms by the first and second order partial derivatives of displacement function. In order to solve the problems of computational complexity and distortion of geometric details, many works~\cite{botsch2005efficient, shi2006fast, zorin1997interactive, kobbelt1998interactive, guskov1999multiresolution, botsch2006deformation} based on multigrid solver or multi-resolution deformation strategy have been proposed one after another. For those works, please refer to \cite{botsch2008linear} for a thorough introduction.} Follow-up works change the form of energy formulation to facilitate the solution and achieve better results. The most widely used is Laplacian-based mesh editing. The well-known As-Rigid-As-Possible (ARAP) energy is also Laplacian based, and has been applied to deformation until most recently~\cite{liu2019cubic}. We will begin with Laplacian based methods, including ARAP and follow-ups, followed by methods using other formulations.

\subsubsection{Laplacian-based Mesh Deformation} \label{sec:lap}
Kobbelt et al.~\cite{kobbelt1998interactive} were the first to propose a multi-resolution Laplacian-based deformation method, which is able to approximately solve constrained mesh optimization in real time. The readers may refer to \cite{sorkine2005laplacian, sorkine2006differential} which gives an early summary of Laplacian-based mesh processing.
Differential coordinates can capture local properties~\cite{alexa2003differential} used in free-form deformation~\cite{sederberg1986free}, and allow a direct detail-preserving reconstruction of the edited mesh by solving a linear least-squares system~\cite{lipman2004differential}.
However, the differential coordinates are defined in a global coordinate system, and thus are not rotation-invariant, so it is necessary to introduce approximated local frames to compensate for some distortions of orientation~\cite{lipman2004differential}.
Laplacian coordinates, as pointed out by \cite{lipman2004differential}, are the simplest form of the differential coordinates.
Given a triangular mesh model, we denote each vertex of the mesh as $\mathbf{v}_i, i=1,\cdots,n$, where $n$ is the number of vertices. The 1-ring neighborhoods of $\mathbf{v}_i$ are denoted as $N(i)$. Then we can define Laplacian coordinates of the vertex $\mathbf{v}_i$ as
\begin{equation}
\mathbf{l}_i = \sum_{j \in N(i)} {w_{ij} (\mathbf{v}_j-\mathbf{v}_i)}
\end{equation}
where $w_{ij}$ is the weight of the edge $\mathbf{e}_{ij}=\mathbf{v}_j-\mathbf{v}_i$. It can be seen that $\mathbf{l}_i$ is a weighted average of position differences between the vertex $\mathbf{v}_i$ and its adjacent vertices, so it describes the local geometry at $\mathbf{v}_i$. By collecting all Laplacian coordinates $\mathbf{l}_i$ and presenting them in the matrix form, it can be written as $\mathbf{l}=\mathbf{LV}$, where $\mathbf{L}$ is a $3n \times 3n$ matrix with elements composed of weights $w_{ij}$. We refer to $\mathbf{L}$ as the Laplacian operator and its elements as the Laplacian coefficients. Sorkine et al.~\cite{sorkine2004laplacian} propose to minimize the differences between Laplacian coordinates before and after deformation to deform the surface models, which can form a sparse linear system and be solved in the least squares sense. Lipman et al.~\cite{lipman2005laplacian} review the above two Laplacian based methods~\cite{lipman2004differential, sorkine2004laplacian} which both preserve shape details when editing mesh models.

Many works have improved Laplacian coordinates or proposed other forms of differential coordinates.
For example, Yu et al.~\cite{yu2004mesh} propose a gradient domain mesh editing method which deforms meshes by interpolating gradient fields derived from user constraints.
Zhou et al.~\cite{zhou2005large} propose a Laplacian coordinate method based on volumetric graph to better preserve the model volume during deformation.
The above two methods require the orientation and local frames of the handles as input. So if the users only move the handles, the orientation and local frames are not changed accordingly, leading to shearing and stretching distortions caused by incompatible handle positions and orientations.
Pyramid coordinates~\cite{sheffer2004pyramid} and iterative dual Laplacian~\cite{au2006dual} are proposed to solve this problem which preserve rotation invariance and can avoid the distortion caused by incompatible rotation and translation of handles. However, such methods cannot handle large-scale rotations.
To deal with the problem that either cannot get the rotation information when the handles are only translated, or cannot handle large-scale rotations, Fu et al.~\cite{fu2007effective} propose to use an affine matrix at each vertex which is linear w.r.t. the vertex position. They further decompose the affine matrix by polar decomposition to extract only rotation and uniform scaling to offset the impact of shearing distortion. 

Most of the previous gradient based editing methods turning the problem into solving a linear system, but not all the constraints can be formulated linearly and also, non-linear constraints are more flexible. Huang et al.~\cite{huang2006subspace} propose a solution framework that can effectively solve the deformation containing nonlinear constraints, such as skeleton constraints for skeleton-based deformation and volume constraints for volume preservation. Those constraints can be transformed to a non-linear energy minimization problem, but the minimization faces the problems of slow convergence and numerical instability. So they build a coarse cage mesh enclosing the original mesh shape, and use mean value coordinates~\cite{ju2005mean} to transfer the energy and constraints to the cage.

\begin{center}
	\includegraphics[width=0.85\linewidth]{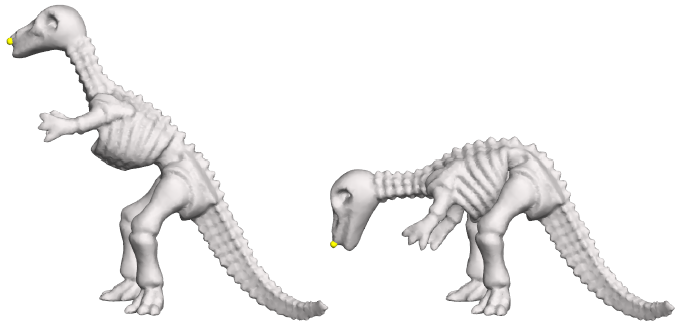}\\
	\vspace{2mm}
    \parbox[c]{8.3cm}{\footnotesize{Fig.3.~}\yyj{Moving} a single position constraint, the result with large deformations can be obtained~\cite{sorkine2007rigid}.}%
\end{center}

\textbf{As-rigid-as-possible (ARAP)} deformation is an important part of Laplacian-based methods.
The principle of as-rigid-as-possible (ARAP) was first applied to shape interpolation~\cite{alexa2000rigid} and the deformation of two-dimensional shapes~\cite{igarashi2005rigid}. Sorkine et al.~\cite{sorkine2007rigid} further propose a 3D surface model deformation method that maintains local rigidity. The method is based on minimizing an ARAP energy, which measures non-rigid distortions in local 1-ring neighborhoods of all vertices. 

We denote the triangle mesh as $S$, and $N(i)$ is the index set of vertices adjacent to vertex $i$. Denote $\mathbf{v}_i \in \mathbb{R}^3$ as the position of the vertex $i$ on the mesh $S$. Also assume that $S$ is to be deformed to $S'$ with the same connectivity and different vertex positions $\mathbf{v}_i'$. 
The overall deformation energy to measure the rigidity of the entire mesh is the sum of the %
distortion energies of each deformation cell $C_i$ (including vertex $i$ and its 1-ring neighbors):
\begin{equation}
\begin{split}
E(S')&=\sum_{i=1}^{n} \bar{w}_i E(C_i, C'_i) \\
&=\sum_{i=1}^{n} \bar{w}_i \sum_{j \in N(i)} w_{ij} {\| (\mathbf{v}'_i-\mathbf{v}'_i) - \mathbf{R}_i(\mathbf{v}_i-\mathbf{v}_j)\|}^2,
\end{split}
\end{equation}

Here, $C_i'$ denotes the deformed cell of $C_i$, $w_{ij}= \frac{1}{2} ( \cot{\alpha_{ij}} + \cot{\beta_{ij}})$ is the cotangent weight, and $\alpha_{ij}$, $\beta_{ij}$ are the angles opposite of the mesh edge $(i,j)$, $\bar{w}_i$ is the cell weight that needs to be pre-determined, which is set to $1$ in \cite{sorkine2007rigid}. We can notice that $E(S')$ depends only on the geometry of $S$ and $S'$, the positions of the vertices $\mathbf{v}_i$ and $\mathbf{v}'_i$. In particular, when the source model $S$ is determined, the only variables in $E(S')$ are the deformed vertex coordinates $\mathbf{v}_i^{'}$. This is because the optimal rotation matrix $\mathbf{R}_i$ is a function of $\mathbf{v}^{'}_i$.

\cite{sorkine2007rigid} takes vertex $i$ and its 1-ring neighbors as a cell, and each cell seeks the best rotation matrix $\mathbf{R}_i$ that satisfies the condition as much as possible. The overlaps between the cells ensure continuous deformation. \yyj{They further formulate an iterative optimization framework that is easy to implement, which readers can refer to \cite{sorkine2007rigid} for details.} The deformation shows the advantages on detail preservation and elastic effect. Given only position constraints, reasonable deformation results can be obtained, as shown in Fig.3.

Following \cite{alexa2000rigid,igarashi2005rigid,sorkine2007rigid}, many ARAP extensions have been developed. Applied to volume deformation, Chao et al.~\cite{chao2010simple} derive another discretization of ARAP energy from the continuous form. The ARAP energy can be further enhanced by smooth rotations~\cite{levi2014smooth}, which can achieve comparable results on surface mesh models to volumetric ARAP~\cite{chao2010simple} on tetrahedral meshes, as shown in Fig.4. Cuno et al.~\cite{cuno20073d} formulate the ARAP deformation with a Moving Least Squares (MLS) approach.
Liu et al.~\cite{liu2011rigid} extend \cite{alexa2000rigid} and propose a new morphing method for surface triangular meshes based on ARAP. Compared to \cite{alexa2000rigid}, their method does not need tetrahedral meshes to represent the shapes, which reduces computation time, and by integrating the translation vector into the energy formulation, eliminates the need for users to specify the fixed vertices when solving the equation.

ARAP deformation has also been extended to make the stiffness of deformation  controllable.
Instead of using 1-ring neighborhoods, Chen et al.~\cite{chen2017rigidity} specify larger neighborhood sizes to better preserve geometric details, and also offer a parameter to adjust physical stiffness.
Qin et al.~\cite{qin2020surface} replace 1-ring neighborhoods with a face-based local cell and specify a stiffness parameter for each local cell to simulate deformation of different materials.
Le et al.~\cite{le2020stiff} extend ARAP deformation to editing man-made models. They improve stiffness of ARAP deformation by introducing an anisotropic material and a membrane model. However, the deformation focuses on local stretching.
Another approach to extending ARAP to anisotropic ARAP was proposed by Colaianni et al.~\cite{colaianni2016anisotropic, colaianni2017anisotropic} that introduces an affine matrix $\mathbf{T}_i$ into the ARAP formulation to enable directional editing:
\begin{equation}
E(S')=\sum_{i=1}^{n} \bar{w}_i \sum_{j \in N(i)} w_{ij} {\| (\mathbf{v}'_i-\mathbf{v}'_i) - \mathbf{T}_i \mathbf{R}_i (\mathbf{v}_i-\mathbf{v}_j)\|}^2
\end{equation}
Different forms of matrix $\mathbf{T}_i$ can realize anisotropic scaling, \yyj{anisotropic shearing or anisotropic rotation}.

\begin{center}
	\includegraphics[width=0.9\linewidth]{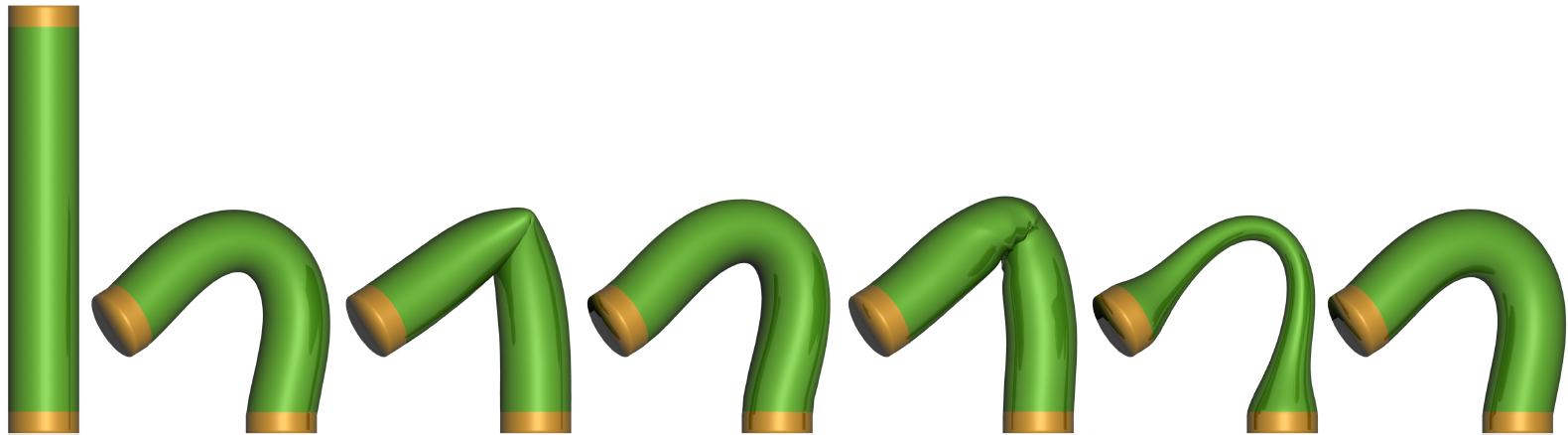}\\
	\vspace{2mm}
    \parbox[c]{8.3cm}{\footnotesize{Fig.4.~}Comparison of SR-ARAP (smooth rotation ARAP) with some other deformation methods~\cite{levi2014smooth}. From left to right: source, PriMo~\cite{botsch2006primo}, ARAP surface~\cite{sorkine2007rigid}, ARAP volume~\cite{chao2010simple}, ARAP volume applied to a tetrahedral stratum, ARAP surface with additional term for a smooth map differential, SR-ARAP~\cite{levi2014smooth}.}%
\end{center}

\begin{center}
	\includegraphics[width=0.9\linewidth]{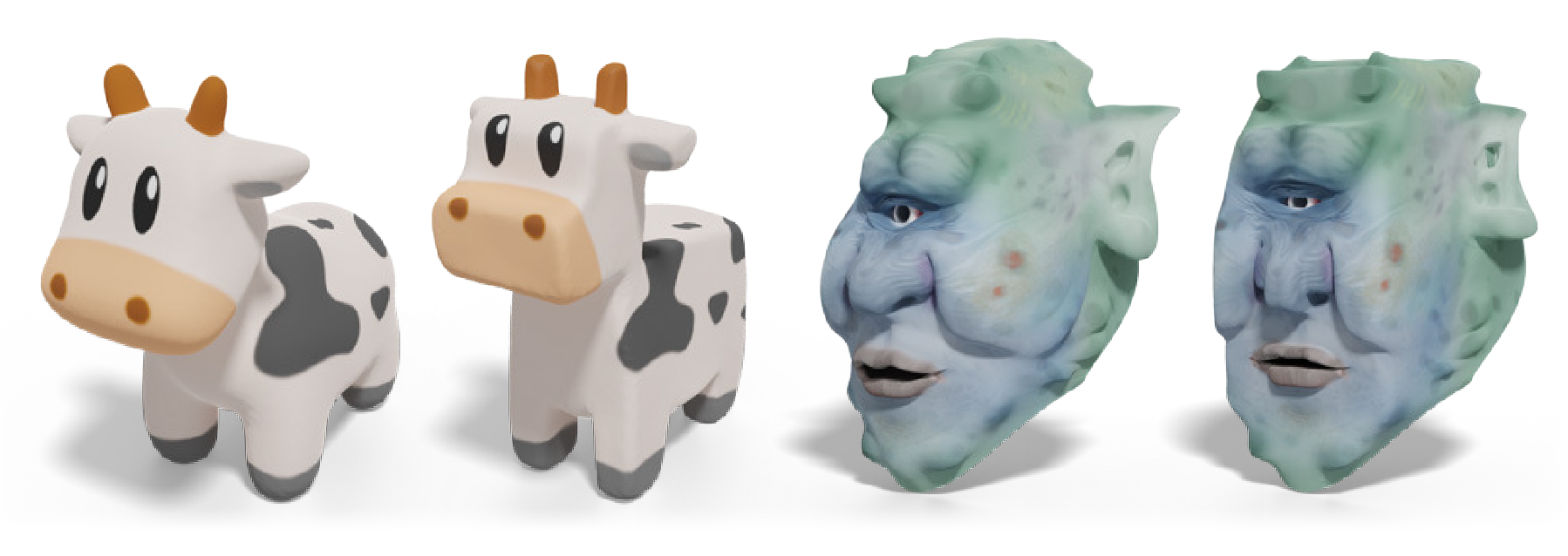}\\
	\vspace{2mm}
    \parbox[c]{8.3cm}{\footnotesize{Fig.5.~}ARAP has been applied to deformation until most recently. Cubic stylization~\cite{liu2019cubic} minimize the ARAP formulation with $l_1$ regularization to achieve locally isometric deformations while preserve texture attributes.}%
\end{center}

ARAP deformation methods are useful, however, they can only achieve interactive rates on coarse meshes~\cite{levi2014smooth}. Some research works investigate acceleration techniques of ARAP.
Borosan et al.~\cite{borosan2010hybrid} combine surface-based deformation with cage-based deformation to perform hybrid mesh editing. The user deforms the simplified version of the input shape using ARAP surface modeling~\cite{sorkine2007rigid}, and the deformation is then propagated to the original shape by precomputed Mean Value Coordinates~\cite{ju2005mean}. Manson et al.~\cite{manson2011hierarchical} also perform ARAP on simplified meshes and 
propose a prototype of hierarchical ARAP. They build coarse meshes using edge contraction, and reverse the edge-collapse process to add details back after deformation on the simplified mesh. Following this acceleration strategy, Liu et al.~\cite{liu2019cubic} achieve cubic stylization of models by minimizing the ARAP energy with an $l_1$ regularization.
Sun et al.~\cite{sun2018bi} also achieve hierarchical ARAP by constructing a bi-harmonic surface to decompose the mesh.
Zollhofer et al.~\cite{zollhofer2012gpu} propose a GPU-based multi-resolution ARAP implementation, which accelerates the computation of ARAP and allows to pose even high-quality meshes consisting of millions of triangles in real-time.
Accelerating the optimization of ARAP has also been addressed in various recent works~\cite{kovalsky2016accelerated, peng2018anderson, rabinovich2017scalable, shtengel2017geometric, zhu2018blended}.

ARAP formulation has also been combined with other deformation methods.
Zhang et al.~\cite{zhang2010skeleton, zhang2011robust} integrate a skeleton into ARAP surface modeling, effectively extending it to volume modeling. They evenly sample points on the skeleton and connect surface vertices \yyj{with the sampled points} to form skeleton edges, which are also considered in an ARAP energy together with the surface edges. In this way, the method can avoid volume loss, which is a common issue for surface-based deformation.
Jacobson et al.~\cite{jacobson2012fast} introduce the ARAP energy into the LBS (Linear Blend Skinning) deformation method to reduce the degrees of freedom that require user specification. They further cluster the vertices based on their Euclidean distances in the \yyj{skinning} weight space, 
and use the same rotation matrix for all the vertices in the same cluster.
Yang et al.~\cite{yang2018biharmonic} propose to combine the ARAP energy with a data-driven energy in deformation transfer. Their method also clusters vertices. However, their clustering is based on the rotation-augmented weight matrix, which is composed of the weight matrix and the ACAP (as-consistent-as-possible) deformation feature~\cite{gao2019sparse} (see Sec.~\ref{sec:ddmeshdeform} for more details). The resulting clusters are more reasonable than using the weight matrix alone.
In addition to the above combinations, the ARAP energy has also been extended for use in other applications such as parametrization~\cite{liu2008local}, data-driven interpolation~\cite{gao2013data}, shape optimization~\cite{bouaziz2012shape}, shape decomposition~\cite{huang2009shape}, mass-spring simulation~\cite{liu2013fast}, image registration~\cite{levi2012d,sykora2009rigid}, image warping~\cite{fadaifard2013image}, and video stabilization~\cite{wang2013spatially}.

\begin{center}
	\includegraphics[width=0.9\linewidth]{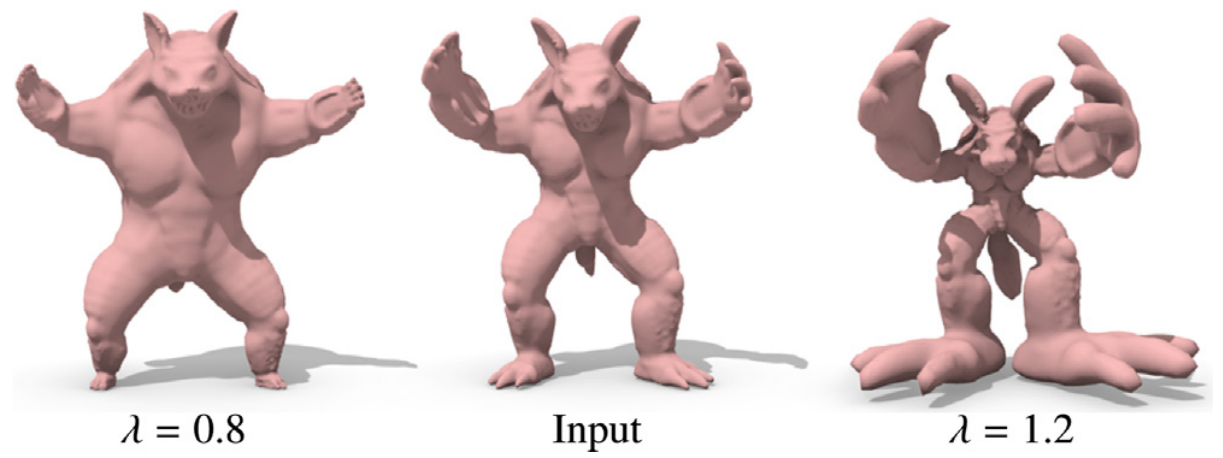}\\
	\vspace{2mm}
    \parbox[c]{8.3cm}{\footnotesize{Fig.6.~}Edit the shape by keeping the mean curvature while changing the Gaussian curvature~\cite{fang2020metric}. As the control parameter $\lambda$ increases, the details are preserved and the main structure is exaggerated.}%
\end{center}

\subsubsection{Other Surface Geometry Properties}
In addition to the Laplacian-based methods for analyzing the local characteristics of the mesh, there are many other geometry-based deformation methods that analyze surface mesh characteristics. For example, curvature is an important attribute of the surface, Crane et al.~\cite{crane2011spin} edit shapes by manipulating the mean curvature and boundary data. The deformation is conformal and has less distortion. 
Fang et al.~\cite{fang2020metric} utilize not only the mean curvature but also Gaussian curvature to perform editing. An example is shown in Fig.6. They also perform conformal surface deformation which preserves local texture features. 

Stretching the mesh may destroy the geometric details.
Alhashim et al.~\cite{alhashim2012detail} propose a shape editing method for stretching, which replicates the geometric details along the stretching direction such that the geometric details are not distorted. They first use a base mesh to represent the general shape of the input, and then use the curve skeleton \yyj{extracted by \cite{au2008skeleton}}
to create a curvilinear grid on the desired stretching region, and project the region onto the grid to form a 2D texture. The user specifies the stretching direction by drawing a 3D curve, and the new geometric details will be synthesized according to the 2D texture.

Liu et al.~\cite{liu2014scale} present a set of scale-invariant measures consisting of triangle angles and edge angles (dihedral angles of the edges) to represent 3D shapes. The representation is unique for a given shape. Moreover, given one edge and 
the orientation of one of the triangles 
containing this edge, the mesh shape can be reconstructed by this representation uniquely. %
The reconstruction is through an iterative process that alternatively solves the face normals and vertex coordinates. 
An ARAP-like formulation is introduced when 
updating the normals, and when solving for the vertex coordinates, the constraints obtained from the user's edited handles are added. The editing process preserves the local details at different scales.

Sparsity has also been widely used in geometry-based mesh deformation. Xu et al.~\cite{xu2015survey} review these methods in geometric modeling and processing that use sparsity, with one section discussing shape deformation based on sparsity. Gao et al.~\cite{gao2012p} introduce general $l_p$ norms to shape deformation, and show that different $p$ values influence the distribution of unavoidable distortions.  Deng et al.~\cite{deng2013exploring} 
explore local modifications of the shape,
and propose to use a mixed $l_2/l_1$ norm regularization which provides more local editing. Different from \cite{gao2012p} that applies sparsity penalty on the error function, they impose it on the displacement vectors.

In addition to the explicit mesh representation, implicit representations such as distance fields or level sets also provide an efficient representation for some editing operations.
Museth et al.~\cite{museth2002level} propose a level set method for surface editing. They define the speed function which describes the velocity at each surface vertex along the surface normal. Different speed functions develop different surface editing operators, such as cut-and-paste operator for copying, removing and merging level set models, and smoothing operator for smoothing the enclosed surface to a predefined curvature value. The method enables easy blending and topological changes of models thanks to the flexibility of implicit representation. 
Eyiyurekli and Breen~\cite{eyiyurekli2017detail} also operate on level set representations and  aim to prevent loss of surface details caused by movements in the direction of the surface normal. Inspired by the idea of multi-resolution deformation, they extract geometric details in advance and store them in the particles on the surface, and then combine the details back when the deformation is completed.

\subsection{Semantic Constraints for Man-made Models} \label{sec:semantic}
Assuming that the input models are all meshes, the simplest way to edit the 3D model is to change the coordinates of the mesh vertices, but this way lacks the necessary constraints and is difficult to produce reasonable results. Therefore, we prefer to use high-level editing methods to edit multiple vertices at same time, such as Free Form Deformation (FFD)~\cite{sederberg1986free,coquillart1990extended}, which we will discuss with cage-based deformation in Sec.~\ref{sec:cage}. 
Although this method is simple and straightforward, the users are required to adjust all parameters manually. Structure is only implicitly imposed by using only a few, low-frequency basis functions. It should be noted that structure is an important indicator in the editing of man-made models. \cite{mitra2014structure} summarizes the method of structure-aware shape processing.

\begin{center}
	\includegraphics[width=0.9\linewidth]{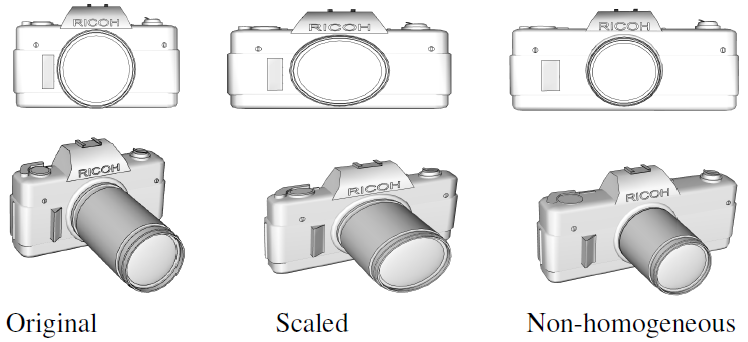}\\
	\vspace{2mm}
    \parbox[c]{8.3cm}{\footnotesize{Fig.7.~}The non-homogeneous resizing results of stretching a camera which preserve structural features~\cite{kraevoy2008non}.}%
\end{center}

\begin{center}
	\includegraphics[width=1.0\linewidth]{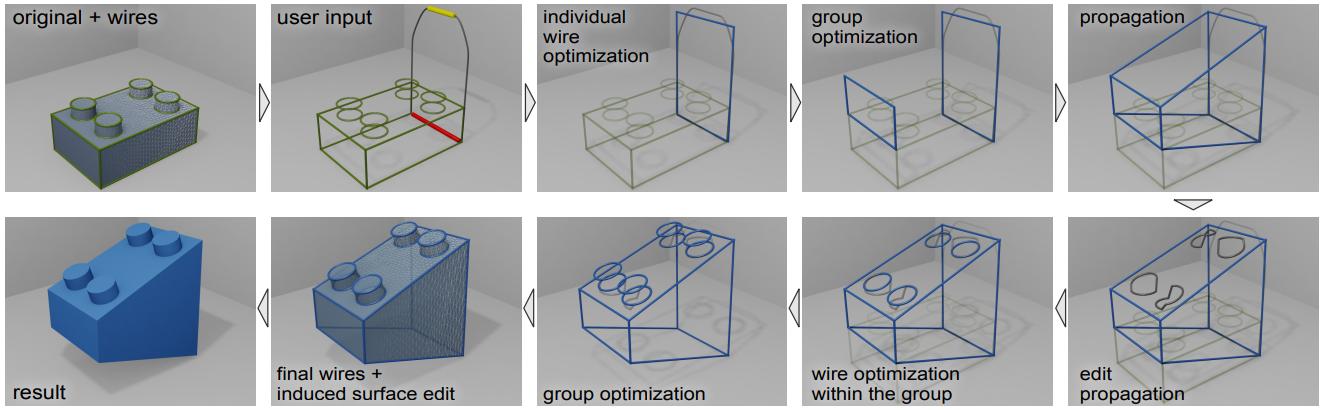}\\
	\vspace{2mm}
    \parbox[c]{8.3cm}{\footnotesize{Fig.8.~}The pipeline of iWIRES~\cite{gal2009iwires}.}%
\end{center}

\subsubsection{Local adaptivity}
Early work on 3D man-made model editing efforts sought to maintain the reasonableness of the 3D shape when scaling 3D models. For example, Kraevoy et al.~\cite{kraevoy2008non} propose to estimate the "vulnerability" of local areas of the shape and adaptively deform the shape. This method prefers axis-aligned stretch when editing, as shown in Fig.7. Xu et al.~\cite{xu2009joint} propose a joint-based shape deformation method, which uses joints to segment a 3D shape into parts, while constraining the relative spatial configuration of adjacent parts. The proposed deformation system edits the models under those joint constraints.

\subsubsection{Global relations}
It is not enough to only consider local adaptivity, so some methods explore the relationship between different parts or features of the whole model, and use this as a constraint to edit the model. Gal et al. propose iWIRES~\cite{gal2009iwires}, which forms an analyze-and-edit paradigm, as shown in Fig.8. Based on the observation that man-made shapes can be abstracted by some special 1D line segments and their relationships, they abstract the 3D shape into a set of curves and adopt simple methods to edit shapes while retaining geometry features. 
Utilizing 1D curves to represent the model structure, Li et al.~\cite{li2010analysis} extract a curve network and additional attributes as prior information to reconstruct the 3D model with detailed and interleaving structures from the scanned point cloud. Those extracted high-level curve can be used as handle to edit reconstructed model.
Following the similar concept of analyze-and-edit, Zheng et al.~\cite{zheng2011component} decompose the model into several meaningful parts, and abstract the parts to simple geometric parametric primitives, which names as component-wise controller. During editing, the user manipulates one of the controllers and the applied change is automatically propagated to other controllers to maintain the structural relations among them, such as symmetry, coplanarity and parallelism. The final model is reconstructed with respect to the modified controllers.
Those controllers also serve as deformation handles for image guided shape editing~\cite{xu2011photo}.
Zhang et al.~\cite{zhang2019real} segment a complex input mesh into several different primitives by clustering, which are depicted by a set of shape parameters and vertices coordinates. In editing procedure, they add several different constraints on these parameters to minimize the target energy function. Optimized parameters are then applied to corresponding primitives to change the shape of input mesh.

Architectural models such as buildings are also important editing targets. These models are highly structured and often have many repetitive patterns, such as windows. 
Based on this observation, Bokeloh et al.~\cite{bokeloh2011pattern} first deform the model under user constraints by as rigid as possible deformation method~\cite{sorkine2007rigid} while maintain continuous patterns. They find the repeated patterns in advance by sliding dockers and measure the stretch to determine insertion or deletion of those discrete repeated patterns after the elastic deformation. 
Also finding those discrete or continuous regular patterns, they~\cite{bokeloh2012algebraic} further build a novel algebraic model of shape regularity and characterize the shape as a collection of those linked translational patterns.
For those irregular architecture models, Lin et al.~\cite{lin2011structure} propose an editing method for resizing them. The users are required to specify the box hierarchy and corresponding attribute, such as replicated, scaled and fixed. Those irregular bounding boxes will be transformed into a set of disjoint sequences automatically. These sequences will be processed in turn. During processing, those user-specified operations are performed on corresponding boxes and their enclosed parts, while the remaining sequences are constrained.
Milliez et al.~\cite{milliez2013mutable} decompose the model into different parts, each of which undergoes elastic deformation. They use several alternative rest states for each elastic part, so the deformation energy is computed by considering a set of those alternative rest shapes. The method further perform the corresponding model editing based on the jigsaw-puzzle-type local replacement mechanism on the user's interactive operations, such as replacement, stretching and shrinking, merging and cutting.
Habbecke and Kobbelt~\cite{habbecke2012linear} linearize the constraints that ensure regional and intuitive control in editing process, making real-time or interactive editing possible.
Texture is an important attribute to show the appearance of the model, but it is not considered in the above methods. Cabral et al.~\cite{cabral2009structure} propose an editing method for textured models which update the texture to maintain the texture features while editing the geometry of the model. They use directional autosimilarity, which measures the ability of a texture region to maintain similarity with itself under slight translation.

\section{Proxy-based Deformation} \label{sec:proxy}
We focus on smoothly interpolating deformation along the surface of 3D models in this section, where we drive a proxy to deform the models.
Organic shape has a hinge structure, in addition to directly editing the vertices on the mesh, binding a skeleton to the shape, and driving the surface deformation through the skeleton is also a popular research direction. We summarize these as \emph{skeleton-based mesh deformation} methods.
There is also extensive research of cage based deformation methods that utilize a set of enclosing cages as proxies, which is not only suitable for organic shape but also man-made models. We summarize these \emph{cage-based deformation} methods in Sec.~\ref{sec:cage}.

\subsection{Skeleton-based Mesh Deformation} \label{sec:skeleton}
Skeleton is one of the shape representations that can describe both the topology and the geometry of the shape~\cite{tagliasacchi20163d}. There are various types of 3D skeletons, we refer the readers to \cite{tagliasacchi20163d} for a thorough survey to the state-of-the-art of various 3D skeletons, while we mainly focus on bone-skeleton used for editing and deformation.

\subsubsection{Skeleton Based Skinning} %
Skeleton-based deformation is most commonly used for the deformation of realistic animated characters. It needs user to bind a skeleton to the shape first, which is termed as the bind time. The user then manipulates the skeleton to deform the shape accordingly, which is the pose time. 
Most methods propagate the handle transformations to the deformation of each surface vertex through the weighted blend of handle transformations.
One of the classical methods that use skeleton to drive the deformation of surface mesh is linear blend skinning (LBS), also known as skeleton subspace deformation (SSD)~\cite{magnenat1988joint}.
Let $ \Omega \subset \mathbb{R}^{2} $ or $ \mathbb{R}^{3} $ denote the volumetric domain enclosed by the given shape $S$. We denote the handles by $H_j \subset \Omega, j = 1, ..., n_h$. In fact, LBS is not limited to skeleton-based deformation. A handle can be a single point, a region, a skeleton bone or a vertex of a cage. Here, we focus on the skeleton bone, and others are easy to generalize. A transformation matrix $\mathbf{T}_j$ require user's specification for each handle $H_j$. Then all vertices $\mathbf{v} \in \Omega $ are deformed by their weighted blends:
\begin{equation}
\mathbf{v}_i^{'} = \sum_{j=1}^{n_h} \mathbf{W}_{ij} \mathbf{T}_j \mathbf{v}_i
\label{equ:LBS}
\end{equation}
where $\mathbf{v}_i^{'}$ is the vertex coordinates after deformation, $\mathbf{v}_i$ is the vertex coordinates before deformation, and $\mathbf{W}_{ij}$ is the skinning weight of handle $H_j$ on vertex $i$.

The linear blend weights $\mathbf{W}$ are crucial to the deformation. Usually, the LBS weights are determined by manual assignment or coming from dataset analysis, which not only takes lots of time and effort, but also produce unnatural deformation results due to lack of smoothness. Bang et al.~\cite{bang2018spline} propose a spline interface for users to edit skinning weights interactively.
Some early works use bone heat~\cite{baran2007automatic} or an improved version, bone glow~\cite{wareham2008bone}, to assign the skinning weights.
Jacobson et al.~\cite{jacobson2011bounded} propose bounded biharmonic weights (BBWs), aiming at enabling users to work freely with the most convenient combination of handle type, making deformation design and control easier. The BBWs produce smooth and intuitive deformation results for any topology of control points, skeletons, and cages. They define the weight vector $\mathbf{W}_j$ of the $j$-th handle (consisting of the control point weights at all vertices) as minimizers of a higher-order shape-aware smoothness functional, namely, the Laplacian energy:
\begin{equation}
\mathop{\arg\min}_{\mathbf{W}_j, j = 1, ..., n_h} \frac{1}{2} \int_{\Omega}^{} {\|\Delta {\mathbf{W}_j} \|}^2 dV
\label{equ:BBW}
\end{equation}
subject to: 
$\mathbf{W}_j|_{H_k} = \delta_{jk}$,
$\sum_{j=1}^{n_h} \mathbf{W}_j(\mathbf{v}) = 1$ and
$0 \leq \mathbf{W}_j(\mathbf{v}) \leq 1, j = 1,...,n_h, \forall \mathbf{v} \in \Omega$,
where $\delta_{jk}$ is the Kronecker function, $\mathbf{v}$ is the mesh vertices. 
It is natural that different control handles do not affect each other. The constraints also guarantee that the deformed shape will not scale and the handles all have positive contributions to the deformation.

For the convenient to solve the Laplacian energy Eq.~\ref{equ:BBW}, 
\cite{jacobson2011bounded} discretizes it using the standard linear FEM Laplacian $\mathbf{M}^{-1}\mathbf{L}$,
where $\mathbf{M}$ is the lumped mass matrix and $\mathbf{L}$ is the symmetric stiffness matrix. After discretizing the continuous integral, we can get:
\begin{equation}
\sum_{j=1}^{n_h} \frac{1}{2} \int_{\Omega}^{} {||\Delta {\mathbf{W}_j} ||}^2 dV 
\approx \frac{1}{2} \sum_{j=1}^{n_h} {\mathbf{W}_j}^T(\mathbf{L}\mathbf{M}^{-1}\mathbf{L})\mathbf{W}_j
\label{equ:discretizing_BBW}
\end{equation}
Through discretization, the minimization of an integral is converted into a quadratic optimization which is easy to compute. The above constraints are all linear equations or inequalities about $\mathbf{W}_j$. Once we know the matrices $\mathbf{M}$ and $\mathbf{L}$ of the given shape, the only thing left is solving quadratic optimization problem under linear constraints. We can observe from Fig.~\ref{fig:bbw} that the BBWs are smooth and local. 

\setcounter{figure}{8}
\begin{figure*}[t]
	\centering
	\includegraphics[width=0.9\linewidth]{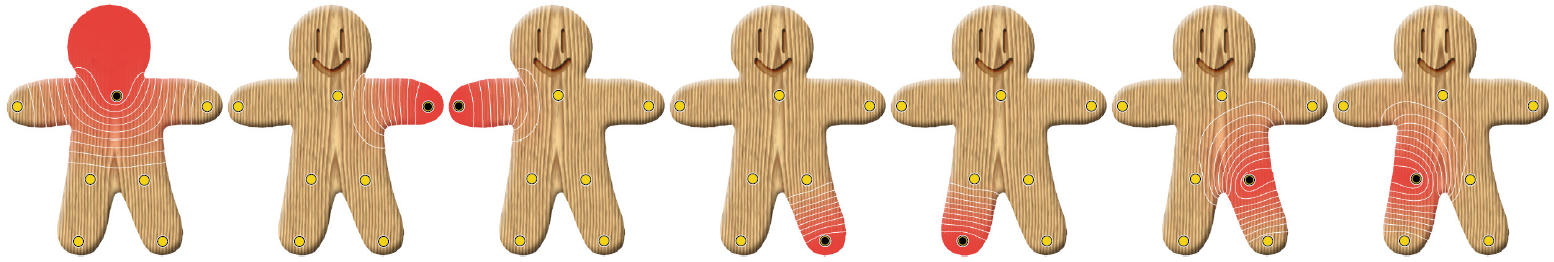}
	\caption{The bounded biharmonic weights~\cite{jacobson2011bounded} are smooth and local.}
	\label{fig:bbw}
\end{figure*}

Directly adding constant bounds to high-order energy leads to more and more oscillation~\cite{jacobson2012smooth}. So Jacobson et al.~\cite{jacobson2012smooth} minimize quadratic energies while avoiding spurious local extrema to wrangle the oscillations.
Exploiting dataset to strengthen the BBWs, Yuan et al.~\cite{yuan2019data} use data-driven, ARAP and sparsity terms to optimize the BBWs. The deformation results using optimized weights can better reflect the deformation principle of the example shapes in the dataset.

The above methods are suitable for manifold meshes. For non-manifold meshes, such as models obtained from 3D modeling software, they are often not watertight or have multiple components. One way of computing skinning weights is to voxelize the model~\cite{dionne2013geodesic, dionne2014geodesic}. The weights are calculated based on the geodesic distance between each voxel lying on a skeleton ``bone'' and all non-exterior voxels. \cite{dionne2013geodesic, dionne2014geodesic} also allow user modify weights interactively when deform the model to test the effect of the modification.

The calculated weights always have an inapplicable area. Eliminating the trouble of assigning weights, Yan et al.~\cite{yan2006skeleton} propose to use skeleton drive the transformation of mesh simplices (triangles in 2D and tetrahedra in 3D) instead of vertices without the need of specification of skinning weights. The vertex connectivity information was directly exploited in their method since simplices include mesh connectivity information.

Although LBS is straightforward, easy to implement and has real-time performance, it can lead to well-known artifacts such as "collapsing elbow" and "candy wrapper". Some methods~\cite{lewis2000pose, mohr2003building, wang2007real, chen2011lattice} have been proposed to address these problems.
Rumman and Fratarcangeli~\cite{abu2015position} first transform the surface mesh to a tetrahedral mesh where LBS is performed, then add stretch constraint, tetrahedral volume constraint and bind constraint to eliminate the artifacts caused by LBS. The constraints are solved by a parallel Position-Based Dynamics schema.
Performing contextual deformation, Weber et al.~\cite{weber2007context} separate surface detail information from skeleton driven pose changes and learn the deformation of skin details from the example characteristic shapes. The editing results can avoid the artifacts of LBS at body elbow.
Shi et al.~\cite{shi2008example} also consider detailed motions (or secondary deformations formally) in skeleton based deformation. They utilize LBS to generate primary deformations and learn those physical behaviors from the example sequences.

In addition to LBS, there are other alternative skinning methods, such as linear combination of dual quaternion or dual quaternion skinning (DQS)~\cite{kavan2007skinning,kavan2008geometric}. However, it suffers more complex vertex processing~\cite{kavan2009automatic}. So they make improvements~\cite{kavan2009automatic} and only use a few samples of nonlinear function (virtual bones) in some key locations, such as joint areas. Other non-linear techniques, such as log-matrix skinning (LMS)~\cite{alexa2002linear, magnenat2004modeling} and spherical blend skinning (SBS)~\cite{kavan2005spherical} also perform volume-preserving deformation, but will suffer bulges near bent joints~\cite{le2016real}. 
Kim and Han~\cite{kim2014bulging} propose some post-processing operations such as modifying vertex coordinates and normals to solve the bulge and distortion problems faced by DQS.

Another choice could be spline skeletons~\cite{forstmann2006fast, forstmann2007deformation, yang2006curve}. They view the bone as a spline and introduce spline deformation to skinning animation replacing the previous transformation matrix guidance. These methods can produce better results but are nonlinear and often fail encountering large rotation deformations. The differential blending method proposed by {\"O}ztireli et al.~\cite{oztireli2013differential} can solve this problem. They use sketch as the interaction tool, and the selected bones will deform to match the strokes drawn by the user.

Jocobson et al.~\cite{jacobson2012fast} combine ARAP~\cite{sorkine2007rigid} with the original LBS formulation, different from some other methods~\cite{wang2002multi, merry2006animation, jacobson2011stretchable} which change the LBS formulation to other forms. All computations of \cite{jacobson2012fast} except for SVD decomposition are linear. When the number of vertex clusters is reasonably selected, real-time deformation can be guaranteed.
Also coping ARAP energy into LBS deformation, Thiery and Eisemann~\cite{thiery2018araplbs} propose a method to generate skinning weights given a 3D mesh and corresponding skeleton. They use a variant of bone heat weights~\cite{baran2007automatic} to initialize the weights and optimize both weights and skeleton joints according to the deformation quality to example shapes.
Li et al.~\cite{li2011skeleton} propose an automatic implicit skinning method which bound the surface onto the skeleton implicitly. The local surface surrounding the joint is used to parameterize the joint position. The deformation is achieved by Laplacian deformation energy with volumetric constraints which prevent those unnatural collapsing at the joints.
Kavan and Sorkine~\cite{kavan2012elasticity} aim to produce visually similar results to physical elastic simulations through skeleton-based skinning method. So they not only propose a new way to calculate the skinning weights but also a new skinning method based on a specific deformer, which they called joint-based deformers. 
Le et al.~\cite{le2016real} propose to impose orthogonal constraints to prevent those artifacts near the joints suffered by LBS, DQS, LMS, and SBS and guarantee real-time performance. However, they need rest pose with skinning weights and bone transformations as inputs.

Artifacts at the joints are often caused by surface self-contact.
Physical-based methods can solve the problem of skin collision well and produce visually plausible deformations, but even after highly optimization~\cite{mcadams2011efficient}, it can only be close to real-time and cannot achieve complete real-time interactive posing.
Vaillant et al.~\cite{vaillant2013implicit} segment the mesh according to the skeleton bones by~\cite{baran2007automatic}, and then approximate each part with an implicit surface utilizing Hermite Radial Basis Functions (HRBF)~\cite{wendland2004scattered, macedo2011hermite}, and at last merge different parts by union or other methods that perform better. They propose to edit the shape through these field functions and geometric skinning methods. The rigid transformations are also applied to the field functions during deformation. The mesh vertices move along the gradient of field function and stop when they reach the original field value or the point where the gradient is discontinuous, so that the surface contacts can be handled well without collision detection. 
Based on \cite{vaillant2013implicit}, Vaillant et al.~\cite{vaillant2014robust} further propose a new family of gradient-based composition operators for combining those implicit surfaces which can deal with surface contacts better. They also derive a tangential relaxation scheme from ARAP~\cite{sorkine2007rigid} to track the iso-surface. The deformation results are better than~\cite{vaillant2013implicit}, especially on extreme character movements.
Teng et al.~\cite{teng2014simulating} apply the subspace simulation of articulated deformable shapes to deal with self-contact situation. They propose a pose-space cubature scheme to resolve the collision without detecting all collision points.

Without the need to input the skeleton or predict the hierarchical structure of the bones, James and Twigg~\cite{james2005skinning} use non-parametric mean shift clustering and least squares method to establish proxy bone transformations and vertex weights to edit and animate the shape.
Yoshizawa et al.~\cite{yoshizawa2007skeleton} propose to extract a skeletal mesh from the dense mesh model. The skeletal mesh is deformed by FFD~\cite{sederberg1986free} and the deformation is back-propagated to the dense model using differential coordinates. A hierarchical framework is used to speed up the process.
Xie et al.~\cite{xie2015agile} propose a shape editing method for personal fabrication applications where the user edit the shape through the constructed skeleton. They observe that most of the editing made by users are local. Based on this fact, they introduce a domain decomposition method that allows the FEM system to re-assemble the sub-matrices only for the local part modified by the user, while the rest remains unchanged, which can avoid unnecessary calculations for fast convergence. 
Xu et al.~\cite{xu2018stress} following \cite{xie2015agile}, also use the skeleton to drive the deformation of the model, and locally update the FEM system. Furthermore, they introduce the multi-grid solvers into the analyzing of the stress distribution. For man-made models, they introduce iWIRES~\cite{gal2009iwires} to preserve the characteristic structure of the model.

\begin{center}
	\includegraphics[width=0.9\linewidth]{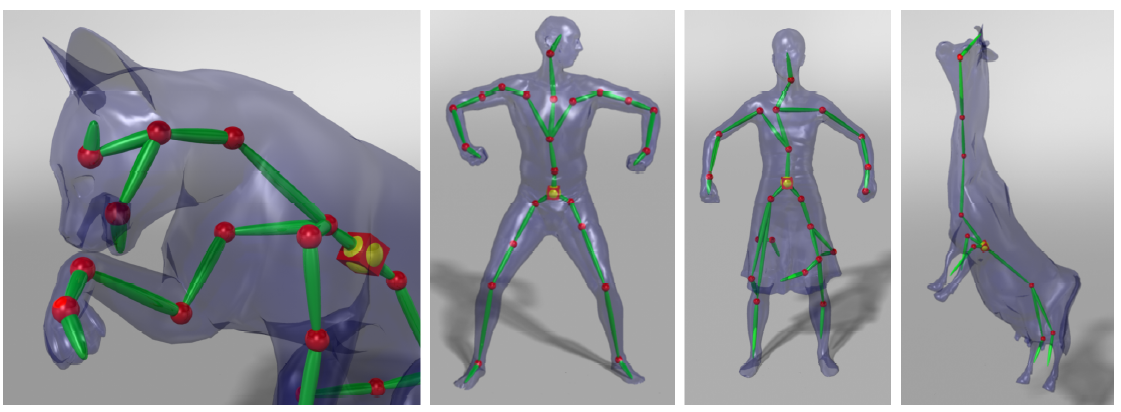}\\
	\vspace{2mm}
    \parbox[c]{8.3cm}{\footnotesize{Fig.10.~}Example-based rigging results of \cite{le2014robust}.}%
\end{center}

\subsubsection{Automatic Rigging}
In addition to studying how to use skeleton to drive shape deformation, another research direction is how to bind a skeleton to the shape. This problem is called \textit{rigging}. In the traditional workflow, this process often need manual specification with the help of professional 3D modeling software. This process usually consists of two steps. The first one is to specify the joint positions and their connections, and the other is to determine the skinning weights which we have mentioned some methods above. There are some works~\cite{au2008skeleton, jiang2013curve, wang2012robust, tagliasacchi2012mean, tagliasacchi2009curve, qin2019mass, livny2010automatic, huang2013l1} that extract skeleton aiming to discover the shape topology, typically called \textit{curve-skeletons}, while we focus on another type of skeletons, called \textit{bone-skeletons}, which can be directly used for editing. For the early work, Baran et al.~\cite{baran2007automatic} propose an automatic method, called Pinocchio, to generate skeleton and the skinning weights from a single shape. They fit a pre-defined skeleton template to the input shape so may fail when the shape structure is different from that of the skeleton. Feng et al.~\cite{feng2015avatar} transfer high quality rigging to input body scan with the help of the SCAPE model~\cite{anguelov2005scape}. However, they only deal with human body shape.
For multi-component characters which is easily accessible on the Internet, Bharaj et al.~\cite{bharaj2012automatically} propose a method to automatically bind skeleton to the character models. The method build contact graph for the components of the input model and exploit graph clustering to obtain the target skeleton with corresponding skinning weights from the input animation skeleton. The mapping from the input skeleton to the target skeleton of the input model are achieved by a novel mapping scheme based on dynamic programming.

Also, the quality of the skeleton extracted using the information from the dataset is better than that extracted from one single shape.
Most works use a set of example poses to extract a hierarchical, rigid skeleton. Schaefer et al.~\cite{schaefer2007example} use clustering to find the rigid bones of the skeleton, and then solve for the vertex skinning weights which are further used to determine the joint positions and their connections. Aguiar et al.~\cite{de2008automatic} bridge the gap between mesh animation and skeleton-based animation. They also first perform clustering to extract rigid bone transformations and then estimate joint motion parameters and appropriate surface skinning weights. Different from former methods that extract skeleton from the examples of same subject, Hasler er al.~\cite{hasler2010learning} estimate a rigid skeleton including skinning weights from examples of different subjects. The skeleton extracted by their method represent either shape variations or pose variations. With the combination of pose skeleton and shape skeleton, user can control them independently. However, Le et al.~\cite{le2014robust} point out that these data-driven methods, on the one hand, use motion driven clustering which does not model the skeleton structure well, so some specific parameter settings are required. On the other hand, the step-by-step process will cause error accumulations. So they adapt skinning decomposition~\cite{le2012smooth} and add soft constraints converting unorganized bone transformations to hierarchical skeleton structure. They over-estimate the number of the skeleton bones during initialization, and exploit iterative framework to automatically prune the redundant bones and update the skinning weights, joint location and bone transformation. The rigging results are shown in Fig.10.

\subsection{Cage-based Deformation} \label{sec:cage}
The cage-based deformation method is very similar to the skeleton-based deformation, but the difference is that the skeleton is generally inside the model, while the cage is generally wrapped outside the model. In essence, they both simplify the structure of the model and provide users with the handle to edit models.
Free Form Deformation (FFD)~\cite{sederberg1986free} is first proposed to produce digital animation. This technique makes it possible to deform 3D shapes smoothly.
Given the lattice vertices $\mathbf{v}_i, i=1,\cdots,n$, we denote the new position of a point inside lattice as $\mathbf{p}'$, and low-frequency basis functions as $\phi_i$, then we can obtain the formulation as follow:
\begin{equation}
	\mathbf{p}' = \sum_{i=1}^{n} {\phi_i(\mathbf{p}) \mathbf{v}_i}
	\label{equ:FFD}
\end{equation}
However, limited by the 3D control lattices, FFD is hard to realize complicated deformations like limb movements, so that it is difficult to depict articulated shapes. Cage based deformation (CBD) is an extension of FFD. The control lattice is replaced by a polyhedral mesh which can better approximate the 3D shape and the deformation formulation is the same as Eq.~\ref{equ:FFD}. 

\subsubsection{Cage Prediction}
The first thing of CBD is cage generation which can be divided into two kinds, automatic and user interactive. Automatic methods are typically completely geometric including mesh simplification~\cite{Cohen1996simplification, Ben2009spatial, Deng2011automatic, Sacht2015nested} and voxelization~\cite{Xian2009automatic, Xian2015efficient}. But these methods tend to produce imperfect cages or sometimes fail. The interactive methods~\cite{Le2017interactive} allow users to add cage vertices to produce better cages for deformation but are more time-consuming. Ju et al.~\cite{ju2008reusable} propose a data-driven method to exploit the cage template dataset created by artists for better cage selection in animation. Savoye et al.~\cite{savoye2010cageik} propose a linear cage estimation method for the target shape given the source shape and corresponding cage, which facilitates cage extraction and animation re-editing work.

\begin{center}
	\includegraphics[width=0.9\linewidth]{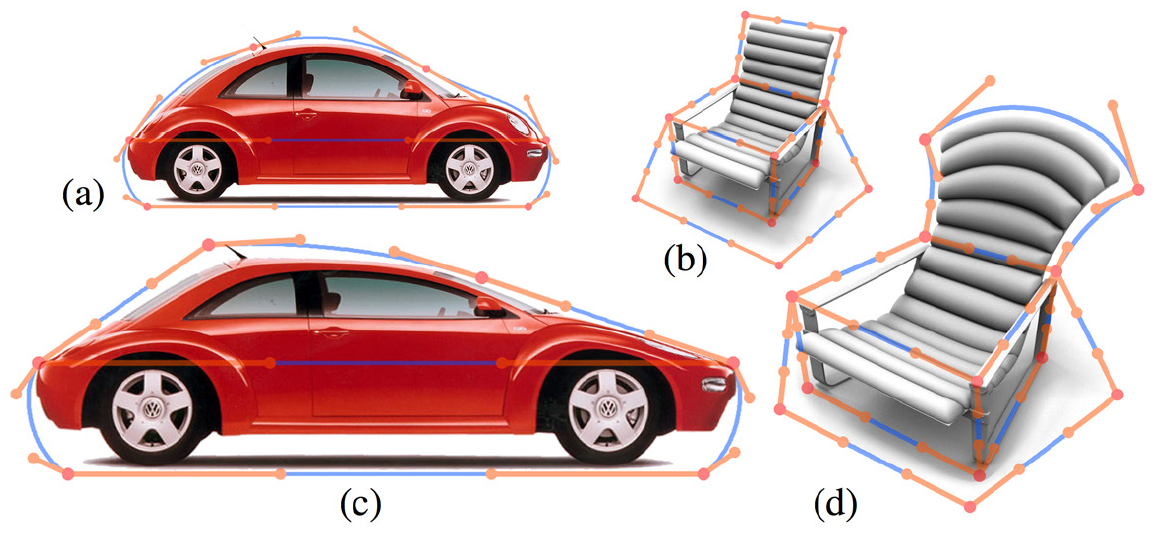}\\
	\vspace{2mm}
    \parbox[c]{8.3cm}{\footnotesize{Fig.11.~}Deformation results using curved edge networks with cubic mean value coordinates~\cite{li2013cubic}. (a) (b) are source models and (c) (d) are edited results.}%
\end{center}

\subsubsection{Blending Weights Generation}
The next thing of cage based deformation is to establish the relationship between the cage and the interior shape. For this purpose, Mean Value Coordinates (MVC) are first introduced by \cite{floater2003mean, floater2005mean} and applied to the deformation for triangular meshes~\cite{ju2005mean}. Hormann et al.~\cite{hormann2006mean} extend MVC to arbitrary polygon meshes. But these coordinates have a main drawback that they could be negative, which will produce unsatisfactory results. 
To avoid the negativeness, Joshi et al.~\cite{joshi2007harmonic} propose Harmonic Coordinates which ensure positive values and produce more local deformations, but the computation is time-consuming. Lipman~\cite{lipman2007gpu} improve MVC to avoid negative values, utilizing GPU visibility render. 
Langer et al.~\cite{langer2006spherical} generalize MVC and vector coordinates~\cite{ju2005geometric} to spherical barycentric coordinates which defined for arbitrary polygonal meshes on a sphere. Those coordinates can also be integrated to existing space-based deformation framework.
Later on, Lipman~\cite{lipman2008green} find that the details of mesh surface are not retained when confronting large-scale deformations. In previous methods like MVC and Harmonic Coordinates, only cage vertex positions are considered. Therefore, he suggests to relate the cage's face normals to the interior vertices and proposes new coordinates called Green Coordinates. 
The Green Coordinates are further extended to complex domain making the deformation better fit the user's input~\cite{weber2009complex}. 
Unlike the original Green Coordinates, which associate the face normals with the vertex positions, the function of the face normals and the function of the vertices in \cite{ben2009variational} are independent, providing a higher degrees of freedom and a larger deformation space.

Yang et al.~\cite{yang2008shape} add global and local stiffness control to the lattice-driven shape deformation. The global stiffness is provided by the width of overlapping lattice cells, and the local stiffness is controlled by the stiffness coefficient. The deformation of the lattice is transferred to the embedded shape by bilinear or trilinear interpolation.
Manson and Schaefer~\cite{manson2010moving} propose moving least squares coordinates which suffer the same problem on boundary edges as MVC and Hermite MVC~\cite{dyken2009transfinite} when used for deforming concave shapes. Weber et al.~\cite{weber2012biharmonic} further propose biharmonic coordinates derived from the solutions to the biharmonic equation. They also present thickness-preserving deformation method which is better than As-Similar-As-Possible (ASAP) and ARAP methods~\cite{igarashi2005rigid}.
In the context of transfinite interpolation, Li et al.~\cite{li2013cubic} propose Cubic Mean Value Coordinates (CMV). Cage-based deformation is essentially a series of interpolation approaches, which interpolate the control vertices of the cage, so CMV can be also used for cage-based shape deformation, as shown in Fig.11. They show shape deformations under the control of cage networks consisting of straight and curved edges.

Most of barycentric coordinates are global, that is, the vertices on the deformed model are determined by the weighted sum of \textit{all} vertices on the cage, which will cause some counter-intuitive deformations, losing good controls for local variations. On the one hand, even for not too many vertices (50-100 vertices) on the cage, the calculation process is time-consuming and may not achieve real-time. On the other hand, since the coordinates are decreasing functions of distance, such as Euclidean distance~\cite{ju2005mean} or geodesic distance~\cite{joshi2007harmonic}, then there are some vertices on the cage that may have little influence on a single vertex of the deformed mesh. So reducing the number of weights is necessary and feasible. Based on these observations, Landreneau et al.~\cite{landreneau2010poisson} propose a Poisson-based weight reduction method which can reduce the number of weights, or saying control points, that affect a single vertex to a user-specified number, while preserving the deformation results the same. They require a certain number (typically 4-6) of example poses in the optimization to achieve better results, and the minimization energy is obtained from Poisson equation solved for the weights by Lagrange multipliers. Their method is also applicable to other deformation methods that require weights, such as skeleton-based deformation methods.
However, imposing the sparsity constraint may obtain the non-optimal solution, which will lead to non-smooth results or even bad approximation results; and sometimes there are exceptional vertices, which are affected by more bones or control points than the preset threshold. Therefore, Le et al.~\cite{le2013two} propose a two-layer blend skinning model that performs lossy weight matrix compression to avoid imposing sparsity constraints. They add virtual bones as an intermediary between the original bones and vertices. They first blend the transformations of the original bones to obtain the transformations of the virtual bones, and then blend up to two virtual bones to obtain the transformation of each vertex. Although mainly dealing with skeleton-based deformation in their paper, their method could also be used in cage-based deformation after combining with an objective function similar to the one in \cite{landreneau2010poisson}.
Similar to enhancing the locality of the deformation, Zhang et al.~\cite{zhang2014local} propose Local Barycentric Coordinates (LBC) for better local deformation. They introduce total variation (TV) originally used in image smoothing and reconstruction~\cite{chambolle2010introduction}, minimizing which under a couple of constraints of partition of unity, reproduction, and non-negativity. The deformation using LBC can realize multi-scale high-quality editing without any other manual specification.
However, LBC has no closed-form expression and must solve a time-consuming optimization problem dealing with dense mesh models, as pointed out by \cite{tao2019fast}. So they propose a new efficient solver for the optimization of LBC.

Some works exceed the limits of single cage and lattice.
Instead of using a polyhedral mesh cage or control lattice, Botsch et al.~\cite{botsch2007adaptive} propose to use small voxels to enclose the model for space-based deformation. They define a nonlinear elastic energy which supports large-scale deformations. However, the discretization may cause some aliasing problems.
Li et al.~\cite{li2010cage} propose a method to directly interpolate points on the mesh without constructing a whole cage for the mesh, instead, they only build an umbrella-like cell interactively on the partial mesh where users are interested. Considering that Green Coordinates can not ensure conformal deformations with an open cage or umbrella-like cell, they also take the local deformation differences of the cage into account.
Replacing the cages with the Interior Radial Basis Functions (IRBF) center points, Levi et al.~\cite{levi2013shape} improve the cage-based deformation methods based on VHM method~\cite{ben2009variational}. The harmonic basis functions are replaced by IRBF which are defined with respect to centers on the surface of the model. They also place a set of spheres inside the model to minimize local distortions by preserving the shapes of the spheres.
Aiming for multi-level detail and high-quality deformations, Garcia et al.~\cite{garcia2013cages} propose a cage-based deformation method based on star-cages instead of a single cage as traditional methods do. The star-cage consisting of multiple cages that offer easier interaction compared to the single cage.
Based on a new representation, sphere-meshes, that can approximate shapes, Thiery et al.~\cite{Thiery2013sphere} use the sphere-mesh hierarchy as a deformation handle to deform shape well. They~\cite{Thiery2016animated} further apply this representation to the approximation of animated mesh sequences and the skinning weights obtained by skinning decomposition can guide pose editing well.

In the traditional FFD or cage-based deformation, after determining the lattice or cage, without re-parameterization, the user can only use the existing handle to deform the shape. Zhang et al.~\cite{zhang2020proxy} propose a control lattice with adjustable topology, which does not need to re-parameterize the relation between lattice and enclosed shape again after changing the lattice. This method uses a tailored T-spline volume to create the lattice and further uses a refinement algorithm to obtain a proxy, which is a simplified version of the lattice and fits the enclosed shape better. The user manipulates the proxy, driving the deformation of the lattice and the deformation is then transferred to the shape. There are fewer vertices on the proxy than the lattice, which is more convenient to manipulate. However, the method is essentially based on volumetric lattice, which is not as flexible as cage-based deformation in large-scale deformation.

\section{Data-based Deformation with Numerical Model} \label{sec:dataset}
With the development of 3D scanning and registration techniques, geometric shape datasets~\cite{anguelov2005scape, Bogo2014faust} are becoming more and more available on the Internet. Analyzing the existing shapes from the shape dataset to provide prior information for deformation becomes an attractive direction. We summarize these as \emph{data-driven mesh deformation} methods. The structural and semantic knowledge of the man-made model can also be obtained by analyzing multiple models. We summarize these as \emph{data-driven analysis for man-made models} methods.

\subsection{Data-driven Mesh Deformation} \label{sec:ddmeshdeform}
Aforementioned geometry-based methods have some essential weaknesses. On the one hand, they are prone to producing unreasonable deformation results when the user's interaction is insufficient. On the other hand, they have high requirements for meshes
and different models tend to require different parameter settings. To address these limitations, data-driven methods exploit plausible deformations from shape datasets and can produce more natural deformation results without manual selection of parameters or a large amount of user constraints. An important pioneering work is Mesh based Inverse Kinematics (MeshIK)~\cite{sumner2005mesh}, based on which a series of works has been proposed to improve or extend the method. We will group these methods according to different deformation representations of shapes. Another active research area is not limited to editing model pose, but also considering shape.

\subsubsection{Blend Mesh Representation}
In the data-driven deformation of the mesh model, deformation representation is important for representing the model. Euclidean coordinates are the most straightforward way to represent the model, but there are obvious limitations on rotation.

\textbf{Gradient-based representation.}
Deformation gradient is a straightforward gradient-based representation, which is defined as the affine transformation that optimally describes the mapping of local neighborhoods (triangles or one-ring neighbors) from the source mesh to the target mesh.
Sumner and Popovi{\'c}~\cite{sumner2004deformation} use deformation gradients to transfer the deformations between two mesh shapes.
Further, Sumner et al.~\cite{sumner2005mesh} propose MeshIK, a method based on principal component analysis (PCA) to analyze the shape dataset and uses the weighted combination of deformation gradients to edit shapes.
MeshIK is used to produce stylized surface deformation and in analogy to traditional skeleton-based inverse kinematics for posing skeletons, and hence the name of MeshIK. %
Each example shape is represented using a feature vector, containing deformation gradients of triangles describing deformation from a reference model. The deformation gradient has a good property that it is a linear function of the mesh vertices. They further decompose the deformation gradient $\mathbf{T}_{ij}$ for the $j$-th triangle in the $i$-th shape into rotation and scaling/shear components using polar factorization $\mathbf{T}_{ij}=\mathbf{R}_{ij} \mathbf{S}_{ij}$.
The rotation is not linear so if it needs to be interpolated linearly, one can map the rotation from 3D rotations $\mathbf{SO}(3)$ to $so(3)$ of skew symmetric $3 \times 3$ matrices~\cite{murray1994mathematical}. The mapping uses the matrix logarithm and can be reversed by the matrix exponential~\cite{murray1994mathematical}. Then the nonlinear span of the deformation gradient for the $j$-th triangle 
given $m$ example meshes has the following formulation:
\begin{equation}
	\mathbf{T}_j(\mathbf{w})=\exp(\sum_{i=1}^{m} {w_i \log(\mathbf{R}_{ij})} ) \sum_{i=1}^{m} {w_i \mathbf{S}_{ij}}.
\end{equation}
This constitutes the nonlinear feature space, where $w_i$ is the combination weight for the deformation gradient from the $i$-th shape. As shown in Fig.12, given different example models, the editing will produce different results.

Der et al.~\cite{der2006inverse} propose a reduced model for inverse kinematics which is faster than MeshIK~\cite{sumner2005mesh}. They cluster the vertices according to the influence of the control parameters, and replace the same cluster of vertices with a proxy vertex located at the weighted centroid of the cluster. The method takes advantage of the reduced complexity of deformation proxies, \yyj{not relying on geometric complexity} 
to interactively edit even extremely detailed geometry models.
Wampler~\cite{wampler2016fast} exploits the ARAP energy~\cite{sorkine2007rigid} for interpolation between a set of example shapes. The method allows spatially localized interpolation which has more natural transitions. However, it also suffers from the problem of potential non-local editing and 
requiring to solve a complicated system of equations when a large number of examples are given.

\begin{center}
	\includegraphics[width=0.9\linewidth]{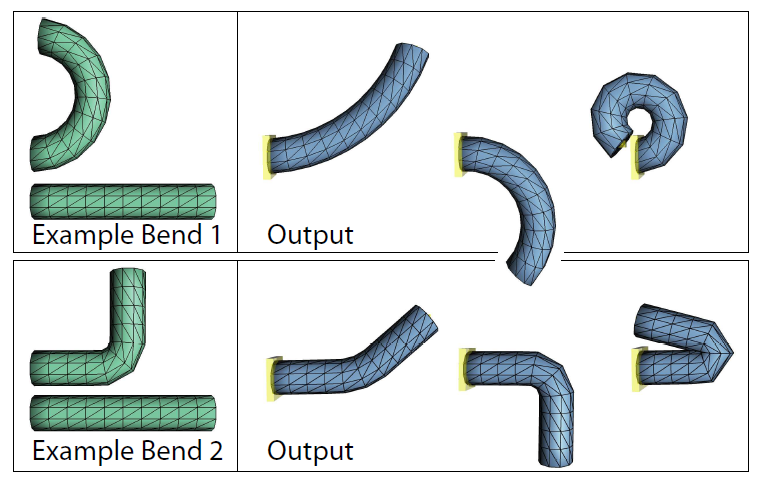}\\
	\vspace{2mm}
    \parbox[c]{8.3cm}{\footnotesize{Fig.12.~}Given different examples, MeshIK~\cite{sumner2005mesh} will produce different deformation results.}%
\end{center}

MeshIK cannot deal with large-scale deformations where rotations are larger than $180^{\circ}$. 
Gao et al.~\cite{gao2019sparse} propose a shape editing method based on ACAP (as-consistent-as-possible) deformation features to address this problem. The rotation at each vertex can be represented using an axis-angle representation. However, the direction of the axis (one of the two opposite directions) and the rotation angle (with multiples of $2\pi$ added) are ambiguous.
They propose an integer optimization strategy to eliminate the ambiguities, so the proposed feature can express rotations greater than $180^{\circ}$.
The method further introduces sparsity constraints into model editing that utilizes the prior information from the model dataset to automatically select a smaller number of basis deformations. It also supports multi-scale editing with high efficiency, as shown in Fig.13.

\begin{center}
	\centering
	\includegraphics[width=0.9\linewidth]{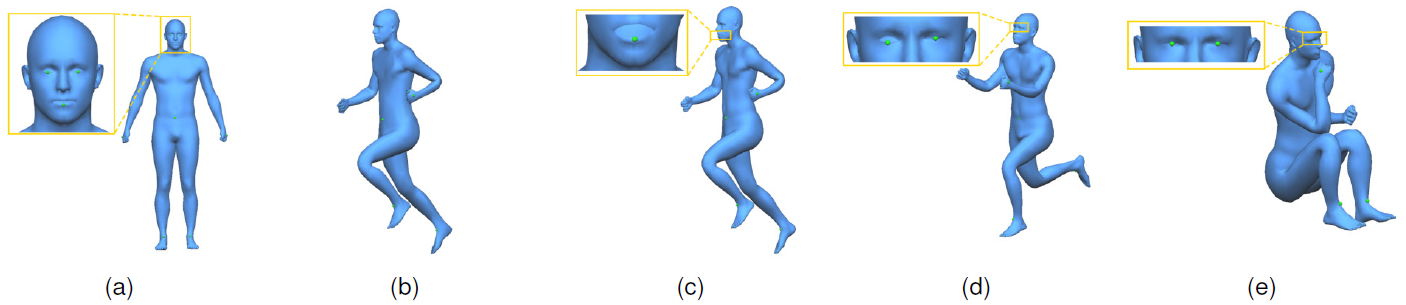}\\
	\vspace{2mm}
    \parbox[c]{8.3cm}{\footnotesize{Fig.13.~}Using ACAP features~\cite{gao2019sparse} along with sparsity constraints enables multi-scale editing. (a) is the reference model. (b) is the deformation result with the simplified mesh. (c)-(e) are the deformed results on the high resolution mesh with both facial and body deformation. Their method automatically selects suitable basis modes for both small-scale facial expression editing and large-scale pose editing.}%
\end{center}

\textbf{Rotation-invariant representations.} 
Another direction of research to tackle rotation ambiguities is to develop rotation-invariant representations.
Lipman et al.~\cite{lipman2005linear} locally define linear rotation-invariant (LRI) coordinates at each vertex which consist of two discrete forms. The discrete form coefficients w.r.t. orientation %
can be used to represent the mesh that facilitate detail-preserving surface editing and shape interpolation.
Changing the definition domain from one-ring
neighborhoods of the vertices to mesh patches, Baran et al.~\cite{baran2009semantic} propose patch-based rotation-invariant coordinates, which solve the noise sensitivity problem of the original LRI~\cite{lipman2005linear} and accelerate the shape reconstruction. They use patch-based LRI coordinates to project the shape into the shape space and transfer semantic deformations to the target shape.
The patch-based LRI representation is further used in data-driven shape interpolation and morphing~\cite{gao2017data} which provide an interface for users to intuitively edit the morphing results.
Kircher and Garland~\cite{kircher2008free} propose a differential rotation-invariant surface representation for surface editing. The second-order differences are both rotation-invariant and translation-invariant. The editing can be operated both in time and space.
Winkler et al.~\cite{winkler2010multi} use the edge lengths and dihedral angles as a representation for multi-scale shape interpolation. Their method supports input settings for more than two shapes.
Further, Fr{\"o}hlich and Botsch~\cite{frohlich2011example} propose to use edge lengths and dihedral angles to represent shape deformation. However, since edge lengths cannot be negative, the method cannot handle extrapolation deformation well.
Gao et al.~\cite{gao2016efficient} propose a data-driven shape editing method based on a novel rotation-invariant representation named RIMD (Rotation Invariant Mesh Difference). They decompose the deformation gradient into rotation and scaling/shear matrices, and combine the logarithm of the rotation difference of each edge and the scaling/shear matrix of each vertex to represent the shape. As shown in Fig.14, the rotation difference cancels out global rotations, making it a rotation invariant representation and thus it can  handle large-scale deformations. However, when applied for data-driven deformation, it uses global principal components extracted from example shapes, so it is difficult to perform local editing. Besides, the derivatives are calculated in a numerical way, which restricts  the editing efficiency.

\begin{center}
	\includegraphics[width=0.9\linewidth]{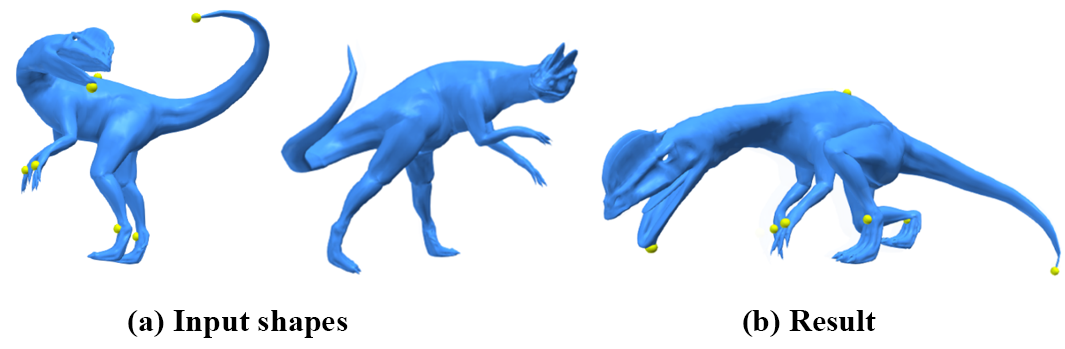}\\
	\vspace{2mm}
    \parbox[c]{8.3cm}{\footnotesize{Fig.14.~}RIMD features can handle large-scale deformations~\cite{gao2016efficient}.}%
\end{center}

Generalized to deforming mesh sequences, Xu et al.~\cite{xu2007gradient} propose a keyframe based mesh editing method. Once the constraints are specified by users or induced from the environment, the frames with those constraints become keyframes. And the constraints and deformations will be propagated to the whole mesh sequence.
Instead of directly editing the input representation of the model, Sumner et al.~\cite{sumner2007embedded} propose to embed the model into a deformation graph which is built by uniformly sampling on the model surface. The graph node $j$ is associated with an affine transformation $\mathbf{R}_j$ and a translation vector $\mathbf{t}_j$ which can map the point $\mathbf{p}$ to a new position by $\mathbf{p}'= \mathbf{R}_j (\mathbf{p}-\mathbf{g}_j) + \mathbf{g}_j + \mathbf{t}_j$,
where $\mathbf{g}_j$ is the position of the graph node. Assuming there are $m$ graph nodes, then the final deformed position $\mathbf{v}'$ of the model vertex $\mathbf{v}$ will be determined by the weighted sum of all influences
\begin{equation}
\mathbf{v}' = \sum_{j=1}^{m} {w_j(\mathbf{v}_i) [\mathbf{R}_j(\mathbf{v}_i-\mathbf{g}_j)+ \mathbf{g}_j + \mathbf{t}_j]}
\end{equation}  
In addition to editing the mesh models, this method can also perform particle simulation, but the disadvantage is that the local details cannot be edited.

\textbf{Deformation components.}
Given a dataset, we can extract the deformation components of the shape and manipulate those basis to achieve the purpose of editing the shape. Early work~\cite{alexa2000representing} employ principal component analysis (PCA) to extract the deformation components, but the extracted components are global, which are not convenient for users to manipulate directly. Therefore, in combination with sparsity, a series of works propose the extraction of sparse deformation components.
For the first work, Neumann et al.~\cite{neumann2013sparse} propose to decompose the animated mesh sequences into sparse localized deformation components (SPLOCS). Those components are spatially localized basis which capture semantic deformations. The user can edit the shape by manipulating those components. However, they operate on vertex coordinates, which are translation and rotation sensitive and thus cannot handle large rotations. Huang et al.~\cite{huang2014sparse} use the deformation gradient to represent the shape, and decompose the deformation gradient into rotation and scale by polar decomposition, and finally use SPLOCS on those vector representations. But this method still cannot handle rotations larger than $180^{\circ}$. 
Bernard et al.~\cite{bernard2016linear} also aim to find local support deformation components from the example shapes in the dataset. They use matrix factorisation with sparsity and graph-based regularisation terms accounting for smoothness to automatically select the position and size of the local support component. 
Adopting rotation-invariant representation, Wang et al.~\cite{wang2017articulated} extend~\cite{neumann2013sparse} using the shape representation of edge lengths and dihedral angles. However, the problem that the extrapolation may fail due to the edge length cannot be negative still exists, and the insensitivity to scale will lead to lack of robustness to noise.
Edge lengths and dihedral angles representation is also used in \cite{liu2019discrete}, which analyzes the edge lengths vectors and the dihedral angles vectors respectively to extract the adaptive sparse deformation components. Then, by adapting \cite{frohlich2011example}, the method allows users to directly edit vertices and produces deformation results under the guidance of components.
Based on Nonlinear Rotation-Invariant Coordinates (NRIC)~\cite{wang2012linear, sassen2020geometric}, Sassen et al.~\cite{sassen2020nonlinear} combine the advantages of principal geodesic analysis~\cite{heeren2018principal} and SPLOCS~\cite{neumann2013sparse} and propose Sparse Principal Geodesic Analysis (SPGA) on the Riemannian manifold of discrete shells.

\subsubsection{Blend Shape and Pose}
This series of methods model the human body through several parameters (often related to shape and pose of the body), and the editing of the human body can be achieved by different parameter inputs.
One of the pioneer work and also one of the most successful work, SCAPE~\cite{anguelov2005scape} uses the deformations of the triangular faces to represent the body shape and pose, separately. The follow-up work, Skinned Multi-Person Linear model (SMPL)~\cite{loper2015smpl}, decomposes body shape into identity-dependent shape and non-rigid pose-dependent shape with vertex-based skinning approach, such as LBS and DQBS.
Given shape parameters $\mathbf{\beta} \in \mathbb{R}^{\|\mathbf{\beta}\|}$ and pose parameters $\mathbf{\theta} \in \mathbb{R}^{\|\mathbf{\theta}\|}$, they propose to represent the neutral mesh $T(\beta,\theta)$ by adding a blend shape function, $B_S(\mathbf{\beta})$, which sculpts the subject identity, and a pose-dependent blend shape function, $B_P(\mathbf{\theta})$ to a mean mesh template $\mathbf{\bar{T}}$,
\begin{equation}
    T(\beta,\theta)=\mathbf{\bar{T}}+B_S(\mathbf{\beta})+B_P(\mathbf{\theta}).
\label{equ:smpl}
\end{equation}
The neutral pose is then deformed by some blend skinning methods,
\begin{equation}
    M(\beta,\theta)=W(T(\beta,\theta),J(\beta),\theta,\mathbf{W}),
\end{equation}
where $W(\cdot)$ represents a standard blend skinning function, and $\mathbf{W}$ is the skinning weights. $J(\beta)$ is a function that determines the joint locations, which transforms rest vertices into rest joint locations.

Although SMPL can model human body well, it lacks modeling of non-rigid dynamic deformations caused by body motions. To model them, Dyna model~\cite{pons2015dyna} proposes to use a second-order auto-regressive model which predicts soft-tissue deformations. Specifically, it represents non-rigid deformation of a body, $\hat{\mathbf{T}}(\beta,\delta)$, by the combination of identity and soft-tissue deformations,
\begin{equation}
    \hat{\mathbf{T}}(\beta,\delta)=\mathbf{S}(\beta)+\mathbf{D}(\delta).
\end{equation}
Further, different from SMPL, Dyna follows the similar idea of SCAPE~\cite{anguelov2005scape}, which describes different human bodies by triangular deformations. Given the edge $\hat{\mathbf{e}}_{i}$ of triangle $i$ in the template mesh, the edge $\mathbf{e}_i$ of triangle $i$ belonging to the mesh at time $t$ to be represented can be represent as,
\begin{align}
    \mathbf{e}_i(\beta,\theta_t,\delta_t)&=\mathbf{R}_i(\theta_t)\hat{\mathbf{T}}_i(\beta,\delta)\mathbf{Q}_i(\theta_t)\hat{\mathbf{e}}_{i} \\
    &=\mathbf{R}_i(\theta_t)(\mathbf{S}_i(\beta)+\mathbf{D}_i(\delta_t))\mathbf{Q}_i(\theta_t)\hat{\mathbf{e}}_{i}
\end{align}
where, $\beta$ and $\theta$ are still body shape coefficients and body pose parameters, respectively. $\mathbf{Q}_i(\theta_t)$ represents pose dependent deformations which are a linear function of $\theta$, $\mathbf{S}_i(\beta)$ represents identity-dependent transformations which are a linear function of $\beta$, $\mathbf{R}_i(\theta_t)$ represents absolute rigid rotations, and $\mathbf{D}_i(\delta_t)$ represents dynamics-dependent deformations which is a linear function of coefficients $\delta_t$. Dynamics deformations are related to body motion, thus, velocities and accelerations. So, the angular velocity and acceleration $(\dot{\theta}_t,\ddot{\theta}_t)$ of body joints and the velocity and acceleration $(v_t,a_t)$ of the root of the body at time $t$ are also the inputs of the model. Let $\hat{\delta}_{t-1}$ and $\hat{\delta}_{t-2}$ be the coefficients representing the history of estimated low-D dynamic deformation, the dynamic control vector of the Dyna model is $\mathbf{x}_t=\{\dot{\theta}_t,\ddot{\theta}_t,v_t,a_t,\hat{\delta}_{t-1},\hat{\delta}_{t-2}\}$ in total. Dynamics deformations also depend on body shape, which is, the shape identity coefficients $\beta$. The dynamics-dependent deformations $\mathbf{D}_i(\delta_t)$ can be further specified as $\mathbf{D}_i(f(\mathbf{x}_t,\beta))$, where $f$ is a function to be learned that maps dynamic control vector $\mathbf{x}_t$ and shape coefficients $\beta$ to the low dimensional representation $\delta_t$ of the dynamics.

SMPL can also be extended to model those dynamic deformations by adding dynamic blend shape function, $B_D(\mathbf{x}_t,\beta)$ to Eq.~\ref{equ:smpl},
\begin{equation}
    T(\beta,\theta_t,\mathbf{x}_t)=\mathbf{\bar{T}}+B_S(\mathbf{\beta})+B_P(\mathbf{\theta_t})+B_D(\mathbf{x}_t,\beta),
\end{equation}
where $B_D(\mathbf{x}_t,\beta)$ also predicts vertex offsets. This model is named as Dynamic SMPL, abbreviated as DMPL.

\begin{center}
	\includegraphics[width=0.9\linewidth]{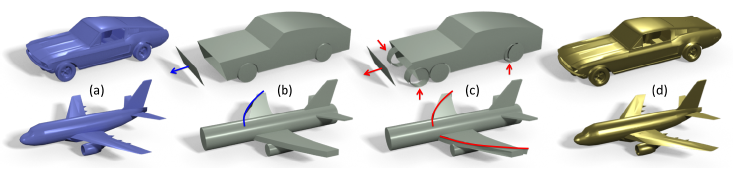}\\
	\vspace{2mm}
    \parbox[c]{8.3cm}{\footnotesize{Fig.15.~}\cite{yumer2014co} supports interactive editing through not only abstract handles but also sketches. (a) Input models. (b) User prescribed deformation (top: translation, bottom: silhouette sketching). (c) Constraints resolved by the system. (d) Final models.}%
\end{center}

\subsection{Data-driven Analysis for Man-made Models} \label{sec:ddanalysis}
Data-driven analysis for man-made model editing is to learn some prior information from a model dataset that contains closely related models, such as belonging to the same category or having the same style. The prior information provide plausible variations of the models and add constraints to user editing which ensure the reasonable results. \cite{xu2016data} reviews the methods of data-driven analysis and processing. 

\subsubsection{Interactive Editing}
Fish et al.~\cite{fish2014meta} propose meta-representation to represent the essence of 3D man-made model dataset. The representation is formulated from the correspondence between model segmented parts, which encodes the arrangement rules of the parts. So it can be viewed as a constraint guiding user editing, where models can maintain their familial traits and performing coupled editing, where several shapes can be collectively deformed by directly manipulating the distributions in the meta-representation. Yumer et al.~\cite{yumer2014co} abstract co-constrained handles for model editing. The handles are obtained from the different segmented parts through a co-abstraction method~\cite{yumer2012co}. The co-constraints are generated by clustering the different planes of the abstracted parts. This method supports interactive editing by not only abstract handles but also sketches, as shown in Fig.15.
Based on this work, Yumer et al.~\cite{yumer2015semantic} further propose a semantic editing method for 3D models, where users can edit 3D models through semantic attributes. This method establishes a continuous mapping between semantic attributes and model geometry through the relative scores of attributes and geometric differences between models. Although the deformation is continuous, this method cannot add and remove certain parts of the model. The above methods all use the dataset of some categories to learn deformation constraints to edit shapes. These methods have been able to take advantage of the information in the shape dataset, and those pairwise parameter constraints work well during shape editing. However, their parameter pairs are in the same kind, and the constraints on the parameter pairs that may be formed by different kinds of parameters is not considered. Based on this, \cite{fu2016structure} use multivariate regression methods to learn the correlation between parameters. The proposed method can perform both structure-preserving and structure-varying shape editing. 
Laga et al.~\cite{laga2017modeling} analyze the pairwise co-variation of the geometry and structure of the part. After the user edits a part of the model, it can automatically find a suitable configuration for the entire model to ensure the plausibility of the edited model.

\begin{center}
	\centering
	\includegraphics[width=0.9\linewidth]{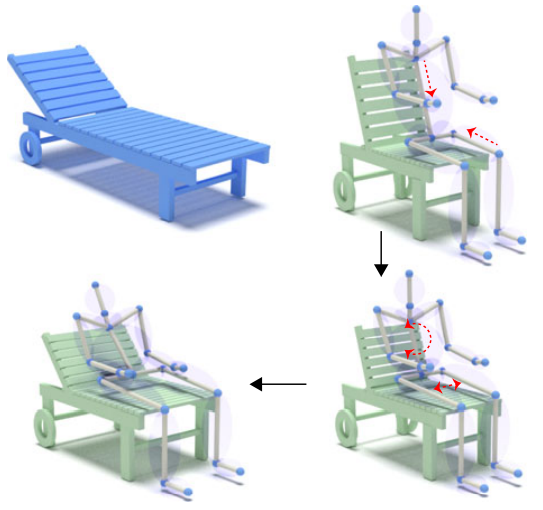}\\
	\vspace{2mm}
    \parbox[c]{8.3cm}{\footnotesize{Fig.16.~}The model changes according to the change of the skeleton~\cite{zheng2015ergonomics}.}%
\end{center}

\subsubsection{Editing for Other Purpose}
From a relatively novel perspective, Zheng et al.~\cite{zheng2015ergonomics} want to change the model to suit for the input human body. As shown in Fig.16, the input is a model with semantic labels, and a spatial relationship graph is used to represent the model, where the graph nodes represent the model components, and the edges of the graph represent the spatial relationships of the components. They first establish the contact constraints between the body skeleton and the model (such as buttocks and chair seats). Then the deformation is an optimization process, and an edit propagation algorithm is designed to deform the model according to these constraints and maintaining the model structure.

Model editing can also be used for other applications. For example, Ovsjanikov et al.~\cite{ovsjanikov2011exploration} explore the shape dataset through the deformations of a template shape which abstract the shape structure using several boxes. Ishimtsev et al.~\cite{ishimtsev2020cad} propose a data-driven mesh deformation method, named CAD-Deform, to fit the retrieved synthetic CAD models to the real 3D scan. The deformation energy ensure smooth deformation and keeping sharp feature, which also include part-to-part mapping term, and nearest-neighbor mapping term. The former match the deformed mesh and the target scan globally, while the latter make them match more accurately when they get close enough.

\section{Neural-based Editing} \label{sec:neural}
In this section, we review attempts on the deformation methods based on deep learning in recent years. Combining with deep learning brings new opportunities and challenges to both organic and man-made shape editing methods.

\setcounter{figure}{16}
\begin{figure*}[t]
	\centering
	\includegraphics[width=0.9\linewidth]{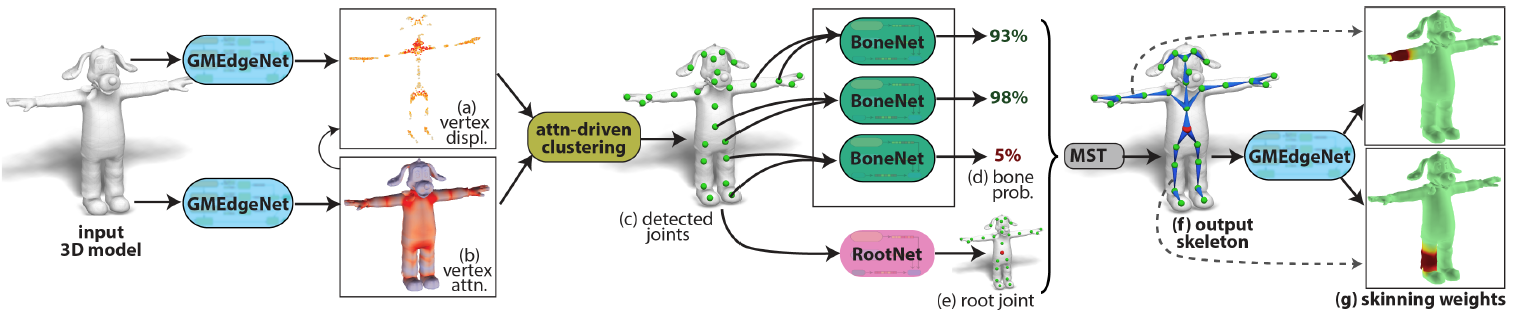}
	\caption{The network architecture of RigNet~\cite{RigNet}.}
	\label{fig:rignet}
\end{figure*}

\subsection{Organic Shape Editing}
With the availability of large human body datasets~\cite{pons2015dyna, AMASS:ICCV:2019}, deep neural networks have also been introduced into the editing of organic shapes.

\subsubsection{Editing via Learning Mesh Deformation}
Tan et al.~\cite{tan2018variational} first propose to use variational autoencoder (VAE) to encode shape deformations. They use RIMD deformation feature~\cite{gao2016efficient} as input and can generate different poses after learning the existing deformations in the dataset. The latent space can be used for shape exploration, which guide the user to find specific shapes they want. However, the network is entirely composed of fully connected layers, which has high memory occupies and thus, cannot handle dense mesh models. To solve this, graph-based convolutions~\cite{defferrard2016convolutional} and mesh pooling~\cite{yuan2020mesh} have been introduced.
At the same time, they~\cite{tan2018mesh} propose a convolutional mesh autoencoder utilizing the locality of the convolution operator and sparsity constraints to extract the local deformation components of the deformable shapes. The deformation components can be used to synthesize new shapes. Also extracting deformation components, Yang et al.~\cite{yang2020multiscale} propose to use multi-level VAEs, which can achieve better results.
Qiao et al.~\cite{qiao2018learning} propose bidirectional LSTM consists of graph convolutions to generate mesh sequences.
As an application of shape deformation and editing, deformation transfer can transfer the user's editing of one shape to another shape. Traditional deformation transfer~\cite{sumner2004deformation, baran2009semantic, Ben2009spatial} need to manually specify several correspondences between source and target shapes. Although Yang et al.~\cite{yang2018biharmonic} propose a method of automatically selecting appropriate key points to transfer the deformation, some candidate points still need to be manually specified. So Gao et al.~\cite{gao2018automatic} first propose a fully automatic shape deformation transfer method. They train SimNet to determine the similarity of two poses of source and target shapes. The proposed VC-GAN combines MeshVAE and CycleGAN~\cite{zhu2017unpaired} to transfer the latent vectors of two input shapes enabling deformation transfer.
Wang et al.~\cite{wang2020neural} represent 3D human meshes by a series of parameters including shape, pose, and vertex order, to perform deformation transfer. The method first encodes these parameters of source shape by a permutation invariant encoder to extract pose feature and then use a style transfer decoder together with target identity mesh as condition to generate the target shape with source pose.

\subsubsection{Performing Mesh Deformation}
Bailey et al.~\cite{bailey2020fast} propose a convolutional neural network for approximating facial deformations which can handle high-frequency deformations. The method separates the process into three parts: a coarse approximation, a refined approximation and an approximation for rigid components of the mesh. The coarse and refined approximations are comprised of two independent convolutional networks by inputting rig parameters. For those segments that only undergo rigid rotations and translations, they are approximated by a faster rigid approximation rather than convolutional networks to improve efficiency. The method also propose a feed-forward neural network to output rig parameters given user-specified control points for inverse kinematics.

The skinning also has some neural-based methods in deformation, skeleton and weights prediction.
As the first work to introduce the neural network in character deformation, Bailey et al.~\cite{bailey2018fast} split skinning deformation into linear and nonlinear portion. The linear portion is approximated by linear skinning method while the nonlinear portion is approximated by a neural network consisting of two fully connected layers. 
Based on the similar idea that decompose the deformation into linear and nonlinear parts, Li et al.~\cite{li2020densegats} propose a graph attention based network to predict the nonlinear effects by inputting mesh graphs and linear deformations, while the linear part is computed with LBS.
Liu et al.~\cite{neuroskinning2019} also propose a neural based skinning method which utilizes the graph convolutions. They first construct a graph using the input 3D mesh with its associate skeleton hierarchy. Each graph node encodes the mesh-skeleton attributes. The graph and node attributes are fed into their graph convolution network to predict the skinning weights.

Almost the same time, Xu et al.~\cite{AnimSkelVolNet} propose to convert an input 3D shape into a set of geometric representations expressed in a volumetric grid. The input representation is processed through a stack of 3D hourglass modules. Each module outputs joint and bone probabilities in the volumetric grid, which are progressively refined by the following module. The final joint and bone probabilities are processed through a Minimum Spanning Tree (MST) algorithm to extract the final skeleton.
They further propose RigNet~\cite{RigNet}, which can predict a skeleton with the skinning weights for the input model with the network shown in Fig.~\ref{fig:rignet}. The method first extracts the geometric feature from the input mesh and predicts candidate joint locations and a attention map indicating the confidence of each candidate joint. After joints are detected, another network learns to determine the root joint and predict whether there is a edge connecting two joints. Finally, a Minimum Spanning Tree algorithm is performed to generate the final skeleton which is sent to another network to predict skinning weights. The method also considers user inputs like how many joints are wanted. The predicted skeleton and skinning weights can be directly used in editing and modeling.
Vesdapunt et al.~\cite{vesdapunt2020jnr} propose joint-based representation for 3D face model which rig semantic joints to the face model. The specified joints add prior information which reduce the demand of large amounts of training data. They also propose an autoencoder network to predict the skinning weights which not only enhance the modeling capacity, but also support users to edit the model.

NNWarp~\cite{luo2018nnwarp} design a heuristic deformation feature vector including geodesic, potential and digression, and warp linear elastic simulations into nonlinear elastic simulations via a DNN prediction to handle a wide range of geometrically complex bodies, which is faster than existing nonlinear method.
Fulton et al.~\cite{fulton2019latent} compress the solid dynamics deformation space into a nonlinear latent space with fewer degrees of freedom through the neural network, while achieving equivalent or even greater simulation fidelity, speed and robustness compared to other model reduction methods. They use the autoencoder architecture and initialize the outer layer with the basis computed by PCA.
Also based on the autoencoder, Santesteban et al.~\cite{Santesteban2020softsmpl} combines the non-linear deformation subspace with a regressor composed of a recurrent architecture GRU which regresses soft-tissue deformations. They propose that the soft-tissue deformations are not only related to shape and pose, but also motion, so the regressor also uses the motion descriptor as input.
NASA~\cite{deng2019neural} and NiLBS~\cite{jeruzalski2020nilbs} condition the implicit field of articulated shapes on the skinning weights, enabling fast shape query without extra acceleration data structures.

\subsection{Man-made Models Editing}
Some large man-made model datasets~\cite{modelnet,shapenet,Thingi10K,ABC} are also available on the Internet, which is the fundamental of some work that try to combine 3D shape editing with neural networks to realize the intelligent editing of 3D shapes.

\subsubsection{Appearance Editing}
Some methods are based on volumetric representation. For example, Yumer et al.~\cite{yumer2016learning} have realized the semantic deformation of 3D shapes by 3D volumetric convolutional network, predicting deformation flow from semantic attributes. However, each semantic attribute is only described by three numbers ($0$, $0.5$, $1.0$, indicating decreasing, keeping the same and increasing respectively), which is lacking in the degree of freedom and controllability of user editing. Liu et al.~\cite{liu2017interactive} realize interactive 3D modeling using adversarial generative networks, as shown in Fig.18. But the edited object is a voxel shape, lacking geometric details, and the resulting shape may not be in line with the user's intentions. As for mesh representation, mesh models generally have inconsistent connectivity. Umetani et al.~\cite{umetani2017exploring} present a parameterization method for efficiently converting a unstructured mesh into a manifold mesh with consistent connectivity using depth information. The parameterization is then fed into an autoencoder, and the plausible deformation space is represented by the latent space of the autoencoder. The users can explore shape variations by directly manipulating the mesh vertices through an interactive interface. Also using autoencoder to optimize on the manifold, DiscoNet~\cite{mehr2019disconet} believes that even if the 3D models belong to the same category, they generally do not lie on a connected manifold. So they propose to use multiple autoencoders (two in their paper) to learn different connected component of the disconnected manifold, without any supervision. Extending the traditional cage based deformation, Wang et al.~\cite{yifan2020neural} propose a neural architecture that predicts source cage and cage offset. The mean value coordinates are computed by a novel MVC layer and a cage-based deformation layer produce the deformed result from the cage offset and mean value coordinates. Also inspired by traditional deformation method, Liu et al.~\cite{liu2021deepmetahandles} propose to use meta-handles, the combinations of control points, as deformation handles and biharmonic coordinates~\cite{wang2015linear} to edit the 3D models. The control points are sampled by farthest point sampling, and meta-handles are predicted by MetaHandleNet. The meta-handles reflect the correlation between the control points. For example, the control points on the two chair armrests should maintain the symmetry of the chair armrests during deformation. At the same time, the plausible deformation range is predicted, and the specific deformation parameters are predicted by DeformNet to deform the source shape to match the target shape. They also propose to use soft rasterizer~\cite{liu2019softrasterizer} and 2D discriminator network to ensure reasonable and realistic deformation.

\begin{center}
	\includegraphics[width=0.9\linewidth]{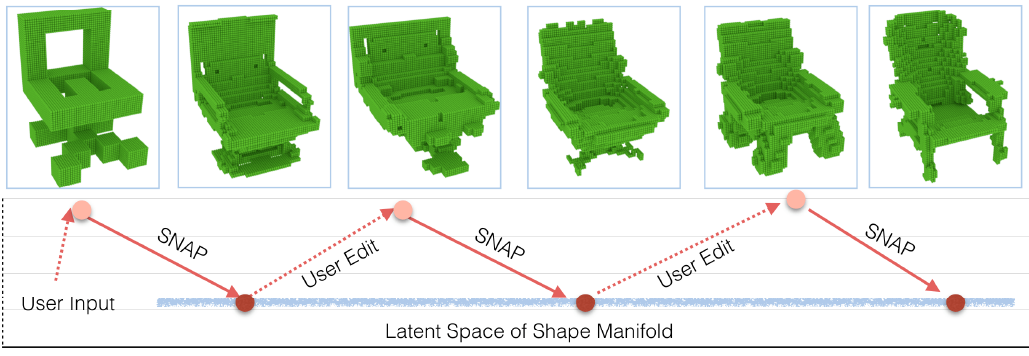}\\
	\vspace{2mm}
    \parbox[c]{8.3cm}{\footnotesize{Fig.18.~}An example of interactive neural editing of man-made 3D models. The user edits the model, and the network maps it to a latent space, and a new model is generated. The result is a voxel model with less geometric details~\cite{liu2017interactive}.}%
\end{center}

DualSDF~\cite{hao2020dualsdf} uses two-level representations to perform interactive shape editing and learn a tightly coupled latent space for the two representations by variational autodecoder (VAD) framework~\cite{zadeh2019variational}. The editing operations are performed on coarse primitive-based representation, and the deformation results are presented as signed distance fields. 
Deng et al.~\cite{deng2020deformed} propose deformed implicit field network(DIF-Net) to represent 3D shape and perform editing. The user freely select one or more 3D points among the surface and specify their new positions. The edited shape is obtained from the latent optimization. The editing also supports adding new structures to the given shape. Also utilizing deformation from a template SDF to represent 3D shapes, Zheng et al.~\cite{zheng2020deep} are able to manipulate shape, but limited to mesh stretching.

Wei et al.~\cite{wei2020learning} propose a encoder-decoder network to edit shapes by editing semantic parameters like height, depth, and width of each semantic part of man-made objects. Their method can be divided into two stages: semantic parameter encoding and deformation transfer. To provide semantic parameters supervision for the encoder, they first generate ground truth semantic parameters for shapes synthesized by bounding boxes of segmented shapes in the real dataset and also edit these corresponding synthetic shapes and get corresponding semantic parameters. After encoding original synthetic shapes and deformed synthetic shapes into the semantic latent space, a decoder use shape-level chamfer distance supervision to reconstruct both original shapes and deformed shapes. At inference time, the network encodes a realistic shape into the parameter space and edit shape parameters on the parameter space and decode the reconstructed synthetic shape and edited synthetic shape. As for deformation transfer, by defining deformation field as the vertex replacement on the decoded synthetic shape, each vertex on the real shape finds $k$ nearest points on the synthetic shape and regards the weighted sum of the displacement of these nearest points as the vertex displacement of the realistic shape. In this way, deformation is transferred from the synthetic shape to the realistic shape. In addition, this method can be easily applied to non-rigid models by changing the definition of the semantic parameters e.g. pose and shape for human bodies.

Sung et al.~\cite{sung2020deformsyncnet} embed shapes into an idealized latent space where points represent shapes and vectors between points represent shape deformations. The deformation vector can be decoded into a deformation action which can be applied to new shape directly.

\subsubsection{Other Forms}
In addition to geometry, structure is also editable. Mo et al.~\cite{mo2020structedit} develop a deep neural network based on structural shape representation StructureNet~\cite{mo2019structurenet} to embed shape differences or deltas into the latent space of VAE, enabling multiple kinds of edits with both geometric and structural modifications.
Representing 3D man-made models as a set of handles, Gadelha et al.~\cite{gadelha2020learning} adopt a two-branch network architecture to generate shape handles. After training the network, users can edit any handle of the handle set, and the back propagation is used to optimize the latent vector to obtain a result that preserve the overall structure.

Reinforcement learning can also be integrated in model editing. For example, Lin et al.~\cite{Lin2020modeling} propose a reinforcement learning based method to edit mesh models. First, the Prim-Agent predicts a sequence of actions to operate the primitives to approximate the target shape given a shape reference and pre-defined primitives. Then the edge loops are added to the output primitives. Second, the Mesh-Agent takes as input the shape reference and the primitive-based representation, and predicts actions to edit the meshes to produce shapes with detailed geometry.

Some methods input some guidance to deform the models. Kurenkov et al.~\cite{Kurenkov2018deform} take an image as input and retrieve a template mesh from the repository, then they deform the template 3D shape to match the input image and preserve the topology of the template shape using Free-Form Deformation.
\yyj{Also retrieving from the given dataset at first, Uy et al.~\cite{uy2021joint} deform the retrieved set of source models to the target image or scan. The retrieval and deformation modules are trained jointly, in an alternating way. The deformation is part-based and structure-aware, predicted by a general MLP which takes encoded target, global and per-part source codes as inputs.}
Wang et al.~\cite{wang20193dn} extracts global features from both the source shape and target input or point cloud. These features are then input to an offset decoder which predicts per-vertex offsets to deform the source to produce a deformed shape similar to the target. Groueix et al.~\cite{groueix2019unsupervised} also perform per-vertex deformation, leveraging not only reconstruction loss, but also cycle-consistency loss.

\setcounter{figure}{18}
\begin{figure*}[t]
	\centering
	\includegraphics[width=0.9\linewidth]{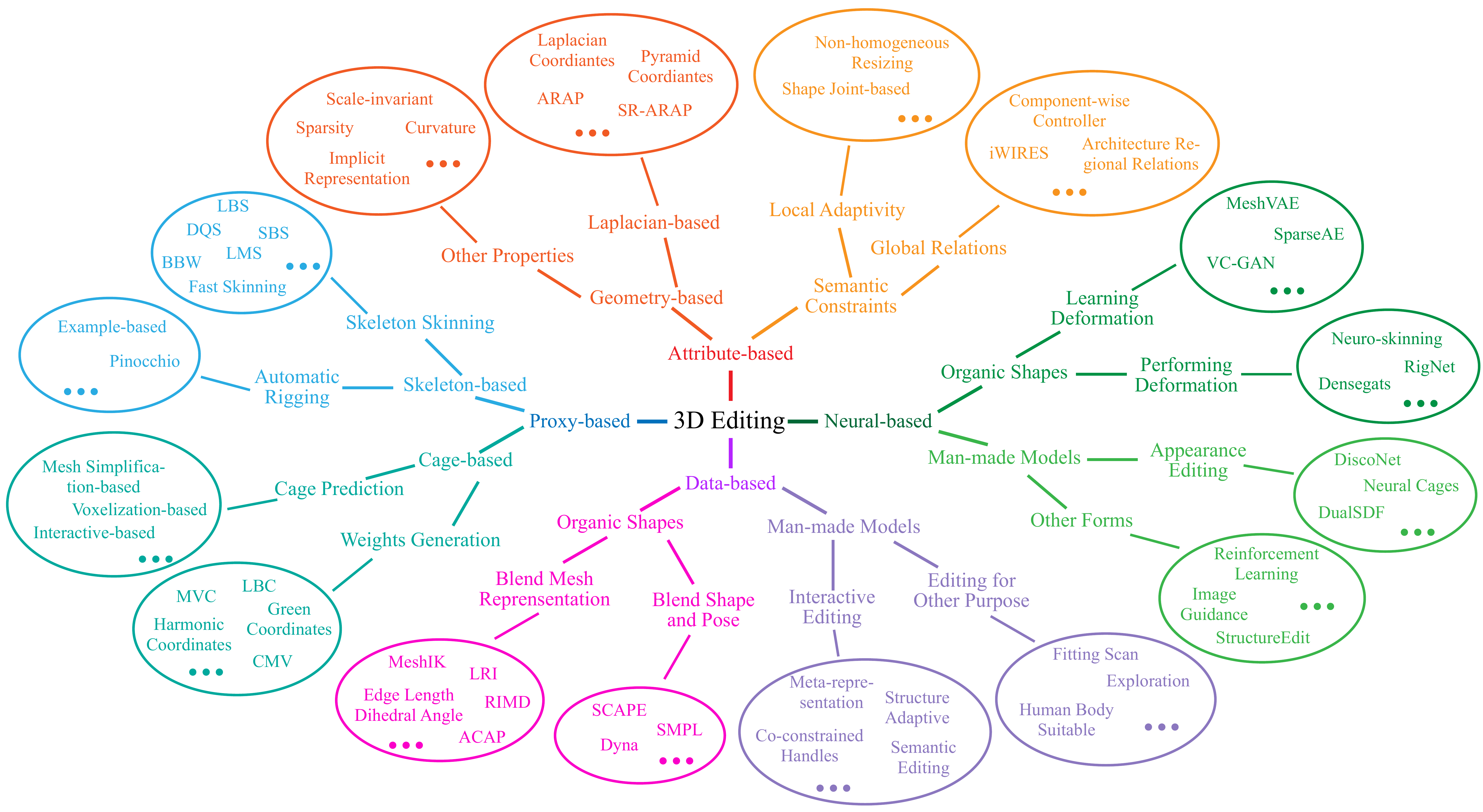}
	\caption{The structure of the survey and some representative works in each subjects.}
	\label{fig:wholemap}
\end{figure*}

\section{Conclusions} \label{sec:conclusion}
In this survey, we have reviewed the history of 3D model editing and the exploration of deep learning based editing methods in recent years.
We divide the editing methods into four subjects based on the data sources. In each subject, we further discuss the respective editing methods around organic shapes and man-made models. The former is generally manifold, and the latter is generally designed by an artist and is non-manifold. \yyj{We show the whole map with some representative works in Fig.~\ref{fig:wholemap}.}

For organic shapes or deformable models, we first discuss classical Laplacian-based methods, especially ARAP~\cite{sorkine2007rigid} and its large amounts of derivative works. In addition to these surface-based deformation methods, there are also deformation methods based on skeleton and cage.
Editing methods learning from dataset take into account the deformation principles of existing models in the dataset, and the deformation results will be more natural. The neural-based methods are also mainly divided into two parts for exploration. On the one hand, they consider the surface meshes and use various deformation representations as input, like traditional data-driven methods. On the other hand, they consider the skinning deformation based on the skeleton to provide intelligent solution strategy for skeleton rigging and weight assignment.

For man-made models, keeping the structure of the model from drastic changes, or maintaining the salient features of man-made models is the most important. Therefore, the editing method of the man-made model will maintain the invariance of the local area, and at the same time analyze the correlation between different parts of the model to limit the editing. This invariance can be obtained by analyzing a single model or a large number of models in the dataset.

Neural-based editing is a promising direction. Although some works have explored neural-based editing methods for 3D models, there are still many directions that can be improved:

\textbf{Organic shapes.} At present, most of the editing methods of organic shapes based on neural networks still use traditional skeleton-based or cage-based skinning methods, while neural network are used for skeleton binding~\cite{RigNet}, cage prediction~\cite{yifan2020neural}, and weight assignment~\cite{neuroskinning2019}. Although there are some methods~\cite{luo2018nnwarp} that explore the direct use of neural networks to predict the displacement of nonlinear deformation, experiments have only been carried out on isotropic materials, and the anisotropic materials need to redesign the framework. Therefore, on the one hand, we still need to design an end-to-end framework which inputs user constraints, such as editing handles and handle displacement, and outputs shape transformation matrix or vertex displacement; on the other hand, we need to study how to relate the selection of the deformation handle with the deformation result, such as optimizing the selection of control points, character rigging, and the prediction of weights according to the deformation results. In the latter, reinforcement learning may be a possible solution, where possible control points are selected by the agent and rewards are given through the deformation results.

\textbf{Man-made models.} The neural editing of man-made model still requires to satisfy both easy-manipulated deformation handle, and representations that can fully show the details of the model. The existing neural-based editing methods either use implicit surfaces~\cite{hao2020dualsdf} or manifold surface~\cite{umetani2017exploring} to approximate the non-manifold model, which will lose part of the model details; or directly use FFD or cage-based editing methods on the original non-manifold model, but the handles are limited, such as semantic vector~\cite{yumer2015semantic} and global deformation through cage~\cite{yifan2020neural}. A good handle can be edge loop~\cite{Lin2020modeling} or coarse primitive-based representation~\cite{hao2020dualsdf}. But how to relate these handles with the non-manifold models still needs lots of work.

\yyj{As a recently widely studied representation, implicit surface can achieve arbitrary resolution theoretically, which is a potential representation in various areas. In addition to further explorations in 3D model editing, \textit{neural morphing}, that is, morphing two shapes using neural networks, and \textit{neural modeling}, that is, modeling 3D models using neural networks, can also be regarded as possible research directions.}

\vspace{2mm}

\bibliographystyle{JCST}
\bibliography{refs}

\label{last-page}
\end{multicols}
\label{last-page}
\end{document}